\documentclass[10pt,conference,letterpaper]{IEEEtran}

\usepackage[OT1]{fontenc} 
\usepackage{times}
\usepackage{float}
\usepackage{algorithm}
\usepackage{makecell}
\usepackage{multirow}
\usepackage{hhline}
\usepackage{algorithmic}
\usepackage{amsmath,bm}
\usepackage[shortlabels]{enumitem}
\usepackage{mathrsfs}

\usepackage{cite}
\usepackage{latexsym}
\usepackage{amsmath}
\usepackage{amsthm}
\usepackage{amssymb}
\usepackage{mathrsfs}
\usepackage[dvips,dvipdf]{graphicx}
\usepackage{subfigure}
\usepackage{multirow}
\usepackage{bbm}
\usepackage{arydshln}
\usepackage{url}
\usepackage{pict2e,picture} % picture
\usepackage{tikz}   % tikz
\usepackage{pifont} % ding

\renewcommand{\a}{\mathbf{a}}

\renewcommand{\b}{\mathbf{b}}

\renewcommand{\d}{\mathbf{d}}

\newcommand{\barf}{\bar{f}}

\newcommand{\I}{\mathbf{I}}

\renewcommand{\u}{\mathbf{u}}

\renewcommand{\v}{\mathbf{v}}
\newcommand{\W}{\mathbf{W}}

\newcommand{\x}{\mathbf{x}}

\newcommand{\y}{\mathbf{y}}
\newcommand{\z}{\mathbf{z}}

\newcommand{\0}{\mathbf{0}}
\newcommand{\1}{\mathbf{1}}

\newcommand{\bepi}{\bm{\epsilon}}

\newtheorem{ex}{Example}
\newtheorem{defn}{Definition}

\newtheorem{assumption}{Assumption}
\newtheorem{myTheo}{Theorem}
\newtheorem{myCor}{Corollary}

\begin{document}
\title{Communication-Efficient Network-Distributed Optimization with Differential-Coded Compressors}
% \author{Paper ID: 1570576431}

\author{Xin Zhang$^{\dag}$ \mbox{\hspace{0.4cm}} Jia Liu$^{\ddag}$ \mbox{\hspace{0.4cm}} Zhengyuan Zhu$^{\dag}$ \mbox{\hspace{0.4cm}} Elizabeth S. Bentley$^{*}$
\\ $^{\dag}$Department of Statistics, Iowa State University
\\ $^{\ddag}$Department of Computer Science, Iowa State University
\\ $^{*}$Air Force Research Laboratory, Information Directorate
%\thanks{
%%{\color{blue} 
%This work has been supported in part by NSF grants ECCS-1818791, CCF-1758736, CNS-1758757, CNS-1446582; ONR grant N00014-17-1-2417, AFRL grant FA8750-18-1-0107, USDA grant 68-7482-17-009.
%DISTRIBUTION STATEMENT A: Approved for Public Release; distribution unlimited 88ABW-2018-4657 on 19 Sep. 2018.
%%}
%}
}

\maketitle

\begin{abstract}
Network-distributed optimization has attracted significant attention in recent years due to its ever-increasing applications.
However, the classic decentralized gradient descent (DGD) algorithm is communication-inefficient for large-scale and high-dimensional network-distributed optimization problems.
To address this challenge, many compressed DGD-based algorithms have been proposed.
However, most of the existing works have high complexity and assume compressors with bounded noise power.
To overcome these limitations, in this paper, we propose a new differential-coded compressed DGD (DC-DGD) algorithm.
The key features of DC-DGD include:
i) DC-DGD works with general SNR-constrained compressors, relaxing the bounded noise power assumption;
ii) The differential-coded design entails the same convergence rate as the original DGD algorithm;
and iii) DC-DGD has the same low-complexity structure as the original DGD due to a {\em self-noise-reduction effect}.
Moreover, the above features inspire us to develop a hybrid compression scheme that offers a systematic mechanism to minimize the communication cost.
Finally, we conduct extensive experiments to verify the efficacy of the proposed DC-DGD and hybrid compressor.
\end{abstract}

\section{Introduction}\label{sec:intro}

Network-distributed optimization, a canonical topic dating back to\cite{tsitsiklis1984problems}, has received significant interests in recent years thanks to its ever-increasing applications, e.g., distributed learning\cite{Boyd11:DistrLearning,Konecny15:FederatedLearning,Nedic17:DistrLrng}, multi-agent systems\cite{Nedic10:Multi_Agent}, resource allocation\cite{Xiao06:RsrcAlloc}, localization\cite{Rabbat04:Loc_Track}, etc.
All these applications involve geographically dispersed datasets that are too big to aggregate due to high communication costs or privacy/security risks, hence necessitating distributed optimization over the network.
%Broadly speaking, in distributed optimization, there are two general models: 1) the {\em master/slave} model (also known as the parameter sever-worker model), where one or multiple parameter servers manage the global decision variables for all worker machines; and 
A notable feature in network-distributed optimization is that there is a {\em lack of shared memory} due to the absence of a dedicated parameter server -- a key component in the hierarchical distributed master/slave architecture.
%Therefore, the global objective function is only partially observable at each node (due to the unavailability of global data).
As a result, every node can only exchange and aggregate information with its local neighbors to reach a {\em consensus} on a global optimal decision.
%Therefore, the standard gradient descent method cannot be directly decentralized in the network-distributed setting, which is in stark contrast to the straightforward implementation under the master/slave model.
%

In the literature, a classic algorithm for solving network-distributed optimization problems is the decentralized gradient descent method (DGD)  proposed by Nedic and Ozdaglar\cite{nedic2009distributed}.
% and spawned a large number of follow-ups ever since.
The enduring popularity DGD lies in its simple {\em gossip-like} structure, which can be easily implemented in networks.
Specifically, in each iteration, the update at each node combines a weighted average of the state information from its local neighbors (obtained by gossiping) and a gradient step based on its own local objective function and state information.
Further, DGD achieves the same convergence rate as the centralized gradient descent method, implying that distributed computation does not sacrifice convergence rate.

However, despite the aforementioned salient features, a major limitation of the DGD method is that it requires full information exchanges of the state variables between nodes.
Hence, the DGD algorithm is {\em communication-inefficient} when solving large-size high-dimensional optimization problems in networks with low-speed communication links.
For example, consider a distributed image regression problem over a satellite network, where each satellite has images of typical resolution $2048 \times 2048$\cite{Garg14:2K2K}. 
In this case, the parameter dimension is $2048\times 2048 \approx 4 \times 10^{6}$ and the communication load per DGD iteration is $134$ MB (32-bit floating-point).
This is problematic for many satellite networks with low-speed RF (radio frequency) links (typically in the range of hundreds Mbps \cite{Shatnaw16:SatCom}). 
To improve DGD's communication efficiency, recent years have seen a line of research based on exchanging {\em compressed} information between nodes (see, e.g., \cite{reisizadeh2018quantized,tang2018communication,zhang2018compressed,reisizadeh2019robust}). 
Specifically, by leveraging various compression techniques (e.g., quantization/rounding\cite{alistarh2017qsgd}, sparsification\cite{wangni2018gradient}), a high-dimensional state space can be represented by a small codebook, hence alleviating the communication load in the network.

However, although progress has been made to various extents, most of the existing works on compressed DGD algorithms suffer from the following key {\em limitations} (see Section~\ref{sec:related} for more in-depth discussions):
1) extra parameter tunings resulted from far more complex algorithmic structures compared to DGD;
2) restricted assumptions on compressors having {\em bounded} compression noise power;
3) convergence speed is slow and sensitive to problem structure;
4) strong i.i.d. (independently identically distributed) distribution assumptions on datasets at different locations, which often do not hold in practice.
In addition, most of the existing works simply treat compressors as ``blackbox operators'' and do not consider how to minimize communication load with specific compression coding scheme designs.
In light of the ever-increasing demand for large-scale network-distributed data analytics, the above limitations motivate us to develop new compression-based algorithms for communication-efficient network-distributed optimization.

The major contribution of this paper is that we propose a {\em differential-coded compression-based} DGD algorithmic framework (DC-DGD), which overcomes the above limitations and offers significant improvements over the existing works.
Moreover, based on the proposed DC-DCD framework, we propose a hybrid compression scheme that integrates gradient sparsification and ternary operators, which enables dynamic communication load minimization.
Our main technical results and their significance are summarized as follows:

\begin{list}{\labelitemi}{\leftmargin=1em \itemindent=-0.5em \itemsep=.2em}

% Talk about the use of SNR-constrained compressors
\item  We propose a new {\em differential-coded} DC-DGD algorithmic framework, where ``differential-coded'' means that the information exchanged between nodes is the differential between two successive iterations of the variables, rather than the variables themselves.
We show that DC-DGD allows us to work with a wide range of general compressors that are only constrained by SNR (signal-to-noise-ratio) and thus could have unbounded noise power.
The use of SNR-constrained compressors {\em relaxes} the commonly adopted assumption on bounded compression noise power in the literature\cite{reisizadeh2018quantized,tang2018communication,zhang2018compressed}.
More specifically, we show that if a compressor's SNR is greater than $(1-\lambda_{N})/(1+\lambda_{N})$, where $\lambda_{N}$ is the smallest eigenvalue of the consensus matrix used in all DGD-type algorithms, then our DC-DGD algorithm achieves the {\em same} $O(1/t)$ convergence rate as the original DGD method. 
%We note that SNR-constrained compressors include a wide range of compression schemes in practice, and are far more practical than the commonly assumed noise-power-constrained compressors in existing works.
%As a result, the proposed DC-DGD framework is much more practical compared to existing works. 

% Talk about simple algorithm structure
\item Not only does the use of SNR-constrained compressors  make our DC-DGD framework more general and practical, it also induces a nice ``{\em self-compression-noise-power-reduction effect}'' that keeps the algorithmic structure of DC-DGD simple.
More specifically, based on a quadratic Lyapunov function of the consensus form of the optimization problem, we show that the accumulated compression noise under DC-DGD shrinks to zero under SNR-constrained compressors and differential-coded information exchange.
Hence, there is {\em no} need to introduce extra mechanisms or parameters to tame the accumulated compression noise for ensuring convergence.
As a result, DC-DGD enjoys the {\em same} low-complexity and efficient convergence rate as the original DGD method.

% Talk about specific coding design.
\item The insights on the relationship between DC-DCD and SNR-constrained compressors further inspires us to develop a hybrid compression scheme that integrates gradient sparsification and ternary operators to obtain {\em controllable} SNR and a high compression ratio simultaneously.
The proposed hybrid compression scheme achieves the best of both worlds through a meticulously designed mechanism to minimize the communication load.
Specifically, under the hybrid compressor, the communication load minimization can be formulated as an integer programming problem.
Based on the special problem structure, we show that the problem can be solved efficiently by a greedy algorithm.
%we develop a greedy algorithm to solve the communication load problem efficiently.

\end{list}

Our results in this paper contribute to the state of the art of theories and algorithm design for communication-efficient network-distributed optimization.
The rest of the paper is organized as follows. 
In Section \ref{sec:related}, we further review related works on the state of the art of compressed DGD-based optimization algorithms.
In Section \ref{Section: DC-DGD}, we first present our DC-DGD algorithm and then analyze its convergence gaurantees.
In Section \ref{Section: Hybrid}, we developed a family of hybrid operators and a greedy algorithm is proposed to choose the optimal hybrid operator.
Numerical results are provided in Section \ref{Section: Numerical}.
We conclude this paper in Section \ref{Section: Conclusion}.

\section{Related Works}\label{sec:related}

As mentioned earlier, compression-based DGD algorithms have received increasing attention in recent years.
In this section, we provide a more in-depth survey on the state of the art in this area to put our work into comparative perspectives.
Broadly speaking, compression-based DGD algorithms can be categorized as follows (some fall into multiple categories):

\smallskip
{\em 1) Uncoded Noise-Power-Constrained Compressed DGD:}
In the literature, most of the early attempts on compressed DGD were focused on noise-power-constrained compressors, which are easier to analyze.
One notable recent work is the QDGD method proposed by Reisizadeh {\em et al.}\cite{reisizadeh2018quantized}.
The main idea of QDGD is to introduce an $\epsilon_t$-scaled aggregation of compressed local copies coupled with an $\epsilon_t$-scaled local gradient step, where $\epsilon_t = O(1/\sqrt{t})$ is an extra diminishing parameter introduced in each iteration $t$ to dampen the noise power.
However, due to the timid gradient step-size $\epsilon_t \alpha$ ($\alpha$ is the original local gradient step-size in DGD), the convergence rate of QDGD is $O(1/t^{1/4})$, which is much slower than the original DGD.
Also, the algorithm is more complex to use than DGD due to the sensitivity in tuning the extra parameter $\epsilon_t$.
Moreover, QDGD was focused on strongly convex cases and it is unclear whether its performance results can be straightforwardly extended to non-convex cases.

\smallskip
{\em 2) Differential-Coded DGD with Noise-Power-Constrained Compressors:}
Another more recently emerging line of research is the differential-coded DGD approach.
For example, in\cite{tang2018communication}, Tang {\em et al.}
proposed the ECD-PSGD algorithm, where extrapolated information is used in each iteration to reduce compression noise.
However, it requires computing an optimized step-size in each iteration, which leads to high per-iteration complexity.
Also, the convergence rate of ECD-PSGD is $O(\log(t)/\sqrt{t})$, which is slower than the original DGD and its stochastic variant.
Another notable example is the ADC-DGD algorithm proposed by Zhang {\em et al.}\cite{zhang2018compressed}, where a $t^{\gamma}$-amplified differential-coded information (with $\gamma > \frac{1}{2}$) is used in each iteration $t$.
It is shown in \cite{zhang2018compressed} that ADC-DGD achieves the same $O(1/t)$ convergence rate as that of the original DGD.
However, ADC-DGD runs the risk of arithmetic overflow due to the asymptotically unbounded $t^{\gamma}$-amplification factor.
This extra $\gamma$-parameter selection of ADC-DGD also makes it complex to use compared to DGD.

\smallskip
{\em 3) Differential-Coded DGD with SNR-Constrained Compressors:}
The most related algorithm to ours is the DCD-PSGD algorithm proposed by Tang {\em et al.} in \cite{tang2018communication}, which is by far the only differential-coded algorithm that can work with SNR-constrained compressors.
Although DCD-PSGD shares the above similarities with us, our DC-DGD algorithm differs from DCD-PSGD in the following key aspects:
i) DCD-PSGD is designed for parallel training, where a key assumption is that the data at each node are i.i.d., which guarantees that the local objectives are identical.
However, our work {\em relaxes} this assumption and allows the local objectives to be non-identically distributed.
ii) The final output of DCD-PSGD is the {\em average} of all nodes in the network, which could be difficult to implement in network-distributed settings.
In contrast, DC-DGD does not require such an averaging at the final output since each node reaches a global optimal consensus.
iii) Although both algorithms work with SNR-constrained compressors,
the SNR constraint of DCD-PSGD is lower bounded by $4(1\!-\!\lambda_N)^2/(1\!-\! |\lambda_N|)^2$, while the SNR lower bound of our DC-DGD is $(1\!-\!\lambda_N)/(1\!+\!\lambda_N)$, where $\lambda_N$ is the smallest eigenvalue of the consensus matrix.
It can be readily verified that our SNR lower bound is much smaller, which implies that our DC-DGD can work with more aggressive compression schemes.
iv) To achieve the best convergence rate, DCD-PSGD requires an optimal step-size determined by a set of complex parameters (cf. step-size ``$\gamma$'' in Theorem~1 and Corollary~2 in \cite{tang2018communication}) and hard to implement in practice. 
In contrast, the step-size selection in our DC-DGD uses simple sublinearly diminishing series and is easy to implement.

\section{Differential-Coded Decentralized Gradient Descent with SNR-Constrained Compressors}
\label{Section: DC-DGD}

In this section, we first present the problem formulation of network-distributed optimization in Section~\ref{sec:formulation}. 
Then, we will present our DC-DGD algorithm in Section~\ref{subsec:algorithm} and its main theoretical results in Section~\ref{sec:main_results}.
Lastly, we provide proof sketches for the main theoretical results in Section~\ref{sec:proof_sketches}.

\subsection{Problem Formulation of Network-Distributed Optimization}\label{sec:formulation}

We use an undirected connected graph $\mathcal{G}=(\mathcal{N,\mathcal{L}})$ to represent a network,
where $\mathcal{N}$ and $\mathcal{L}$ are the sets of nodes and links, respectively, with $|\mathcal{N}| = N$ and $|\mathcal{L}| = E$. 
We let $\x \in \mathbb{R}^{D}$ denote a global decision vector to be optimized.
In network-distributed optimization, we want to distributively solve a network-wide optimization problem:
$\min_{\x \in \mathbb{R}^{D}} f(\x)$, where $f(\x)$ can be decomposed node-wise as follows:
%%%%%%%%%%%%%%%%%%%%%%%%%%%%%%%%%%%%%
\begin{align} \label{eqn_general_problem}
\min_{\x \in \mathbb{R}^{D}} f(\x) = \min_{\x \in \mathbb{R}^{D}} \sum_{i=1}^{N} f_i(\x),
\end{align}
where each local objective function $f_i(\x)$ is only observable to node $i$.
Problem~(\ref{eqn_general_problem}) has many real-world applications.
For example, in the satellite network image regression problem in Section~\ref{sec:intro}, each satellite $i$ distributively collects image data $\{\u_{ij},\v_{ij}, \theta_{ij}\}_{j=1}^{n_i}$, where $\u_{ij}$, $\v_{ij}$, and $\theta_{ij}$ represent the pixels, geographical information, and ground-truth label of the $j$-th image at satellite $i$, respectively, and $N_{i}$ is the size of the local dataset.
Suppose that the regression is based on a linear model with parameters $\x=[\x_{1}^{\top} \, \x_{2}^{\top}]^{\top}$.
Then, the problem can be written as: $\min_{\x} f(\x) \!\triangleq\! \min_{\x} \sum_{i=1}^{N} f_{i}(\x)$,
where $f_{i}(\x) \!\triangleq\! \frac{1}{n_i} \sum_{j=1}^{n_i} ( \theta_{ij} \!-\! \u_{ij}^{\top} \x_{1} \!-\! \v_{ij}^{\top} \x_{2} )^{2}$. 
Note that Problem~(\ref{eqn_general_problem}) can be written as the following equivalent {\em consensus form}:
\vspace{-.05in}
\begin{align} \label{Eq:problem1}
& \text{Minimize} && \hspace{-.5in} \sum_{i=1}^{N} f_i(\x_i) & \\
& \text{subject to} && \hspace{-.5in} \x_i = \x_j, && \hspace{-.5in} \forall (i,j) \in \mathcal{L}. \nonumber
\vspace{-.05in}
\end{align}
where $\x_i\in \mathbb{R}^D$ is the local copy of $\x$ at node $i$. 
The constraints in Problem (\ref{Eq:problem1}) guarantee that the all local copies are equal to each other, hence the name consensus form.

\subsection{The DC-DGD Algorithm} \label{subsec:algorithm}

To facilitate the presentation of our DC-DGD algorithm, we first need to formally define two technical notions.
The first one is the {\em SNR-constrained unbiased stochastic compressors}:

\begin{defn}[SNR-Constrained Stochastic Unbiased Compressor]\label{defn:sdco}
{\em A stochastic compression operator $C(\cdot)$ is said to be unbiased and constrained by an SNR threshold $\eta$ if it satisfies $C(\z) \!=\! \z \!+\!\bepi_{\z}$, with $\mathbb{E}[\bepi_{\z}] \!=\! \0$ and $\mathbb{E}[ \|\bepi_{\z} \|^2] \!\le\! (1/\eta) \|\z\|^2$}, $\forall \z$.
\end{defn}
We can see from Definition~\ref{defn:sdco} that, for a given compressor, $\eta$ is its lowest SNR yielded by its largest compression noise power $\mathbb{E}[\| \bepi_{z} \|^{2}]$.
We note that SNR-constrained stochastic unbiased compressors are much {\em less restricted} than the noise-power-constrained stochastic unbiased compressors previously assumed in the literature (see, e.g., \cite{reisizadeh2018quantized,tang2018communication,zhang2018compressed}), which satisfies $\mathbb{E}[\bepi_{\z}]=\0$ and $\mathbb{E}[\|\bepi_{\z} \|^2] \le \sigma^2$, $\forall~\z$.
That is, the compression noise power is {\em universally upper bounded} by a constant $\sigma^{2}$ regardless of the input signal.
In contrast, the noise power under SNR-constrained compressors could be arbitrarily large as long as it satisfies a certain SNR requirement, hence being more general.
%Hence, SNR-constrained compressors are far more general than noise-power-constrained compressors.
For example, the following are two typical SNR-constrained stochastic unbiased compressors:
\begin{ex}\label{Ex: Sparse}[The Sparsifier Operator~\cite{wangni2018gradient}] 
For any vector ${\z}=[z_1,\cdots,z_d]^{\top} \in \mathbb{R}^d,$ $C({\z})$ outputs a sparse vector with the $i$-th element $[C{(\z)}]_{i}$ following the Bernoulli$(p)$ distribution:
\begin{align*}
\begin{cases}
\mathrm{Pr} ([C{(\z)}]_{i} = \frac{\z_k}{p} ) = p ,\\
\mathrm{Pr} ([C{(\z)}]_{i} = 0 ) = 1-p,
\end{cases}
\end{align*}
where $p \in (0,1]$ is a constant. The operation is unbiased and the SNR is lower bounded by is $p/(1-p)$.
\end{ex}
\begin{ex}\label{Ex: Ternary}[The Ternary Operator~\cite{wen2017terngrad}] 
For any vector ${\z}=[z_1,\cdots,z_d]^{\top} \in \mathbb{R}^d,$ $C({\z})=\|{\z}\|_\infty sign({\z})\circ \b_{\z},$ where $\circ$ is the Hadamard product and $\b_{\z}$ is a random vector with the $i$-th element $[\b_{{\z}}]_{i}$ following the Bernoulli distribution:
\begin{align*}
\begin{cases}
\mathrm{Pr} ([\b_{{\z}}]_{i} = 1 ) = |{{z}}_i|/\|{\z}\|_\infty ,\\
\mathrm{Pr} ([\b_{{\z}}]_{i} = 0 ) = 1-|{{z}}_i|/\|{\z}\|_\infty.
\end{cases}
\end{align*}
The operation is unbiased and the noise power $\mathbb{E}[\| \bepi_{\z} \|^{2}] = \sum_{i=1}^{d} |z_i|(\|{\z}\|_\infty-|z_i|)$ and hence $\eta = \Theta(d)$. 
\end{ex}

Next, we introduce the notion of {\em consensus matrix}, which is denoted as $\W \in \mathbb{R}^{N \times N}$ in this paper.
As will be seen later, the entries $[\W]_{ij}$ in $\W$ define the weight parameters used by each node to perform local information aggregation.
Mathematically, $\W$ satisfies the following properties:
\begin{list}{\labelitemi}{\leftmargin=1em \itemindent=-0.5em \itemsep=.2em}
\item[a)] {\em Doubly Stochastic:} $\sum_{i=1}^{N} [\mathbf{W}]_{ij}=\sum_{j=1}^{N} [\mathbf{W}]_{ij}=1$.
\item[b)] {\em Symmetric:} $[\mathbf{W}]_{ij} = [\W]_{ji}$, $\forall i,j \in \mathcal{N}$. 
\item[c)] {\em Network-Defined Sparsity Pattern:} $[\W]_{ij} > 0$ if $(i,j)\in \mathcal{L}$ and $[\mathbf{W}]_{ij}=0$ otherwise, $\forall i,j \in \mathcal{N}$.
\end{list}
Collectively, properties a) and b) imply that the spectrum of $\W$ (i.e., the set of all eigenvalues) lies in the interval $(-1,1]$ on the real line, with exactly one eigenvalue being equal to 1.
Further, since all eigenvalues are real, they can be sorted as $-1 < \lambda_N(\mathbf{W}) \leq \cdots \leq \lambda_1(\mathbf{W}) = 1$.
For convenience, we define a parameter $\beta \triangleq \max\{|\lambda_2(\mathbf{W})|,|\lambda_N(\mathbf{W})|\} \in (0,1)$, i.e., the second-largest eigenvalue of $\W$ in magnitude.
Simply speaking, the use of the consensus matrix is due to the fact that $(\W \otimes \I_{P}) \x = \x$ {\em if and only} if $\x_i = \x_j$, $(i,j) \in \mathcal{L}$,\cite{nedic2009distributed} where $\x = [\x_{1}^{\top},\ldots,\x_{N}^{\top}]^{\top}$ and $\otimes$ represents the Kronecker product.
Therefore, Problem (\ref{Eq:problem1}) can be reformulated as $\min_{\x \in \mathbb{R}^{D}} \sum_{i=1}^{N} f_{i}(\x_{i})$, $\mathrm{s.t.} \,\, (\W \otimes \I_{P}) \x = \x$, which further leads to the original DGD algorithmic design\cite{nedic2009distributed}.

With the notions of SNR-constrained unbiased stochastic compressors and consensus matrix, we are now in a position to present our DC-DGD algorithmic framework.
To this end, we let $\mathcal{N}_{i} \triangleq \{j \in \mathcal{N}: (i,j) \in \mathcal{L} \}$ denote the set of local neighbors of node $i$.
Then, our DC-DGD is stated as follows:

\medskip
\hrule 
\vspace{.03in}
\noindent \hspace{-.13in} \!\
{\textbf{ Algorithm~1:}} Differential-Coded Compressed Decentralized Gradient Descent Method (DC-DGD)\label{Algorithm: DC_DGD}.
\vspace{.03in}
\hrule
\vspace{0.1in}
\noindent {\textbf{ Initialization:}}
\begin{enumerate} [topsep=1pt, itemsep=-.1ex, leftmargin=.2in]
\item[1.] Set the initial state $\x_{i,0}\!=\! \y_{i,0}=\! \z_{i,0} \!=\!\0$, $\forall i$. 
\item[2.] Let $t\!=\!1$, $\z_{i,1}\!=\!-\alpha_{1} \nabla f_i(\x_{i,0}),$ and $\d_{i,1}\!= \z_{i,1} - \x_{i,0},$ $\forall i$. 
\end{enumerate}

\noindent {\textbf{ Main Loop:}}
\begin{enumerate} [topsep=1pt, itemsep=-.1ex, leftmargin=.2in]
\item[3.] In the $t$-th iteration, each node sends the differential-coded compressed information $C(\d_{i,t})$ to its neighbors, where $C(\cdot)$ is an SNR-constrained stochastic unbiased compressor. 
Meanwhile, upon the reception of all neighbors' information, each node performs the following updates:
%%%%%%%%%%%%%%%%%%%%%%%%%%%%%%%%%%%%
\vspace{-.05in}
\begin{align} \label{Eq: dc_dgd}
&\text{a) Local copy inexact update: } \x_{i,t} \!=\! \x_{i,t-1} \!+\! C(\d_{i,t}). \!\!\! \\
&\text{b) Weighted local aggregation step: } \nonumber\\
& \hspace{.4in} \y_{i,t} = \y_{i,t-1} + \sum\nolimits_{j \in \mathcal{N}_{i}} [\mathbf{W}]_{ij}C(\d_{j,t}). \\
&\text{c) Local gradient step: } \z_{i,t+1} = \y_{i,t} - \alpha_{t} \nabla f_i(\x_{i,t}). \!\!\! \\
&\text{d) Local differential update: } \d_{i,t+1} = \z_{i,t+1} - \x_{i,t}.
\end{align}
\item[4.] Stop if some preferred convergence criterion is met; otherwise, let $t \leftarrow t+1$ and go to Step 3. 
\end{enumerate}
\smallskip
\hrule
\medskip

Several important remarks on the DC-DGD algorithm are in order: 
1) The combined update structure in Steps 3-b) and 3-c) is the {\em same} as the original DGD algorithm, which contains a weighted local aggregation step and a local gradient step.
Notably, DC-DGD only has one parameter: the step-size $\alpha_t$ (same as DGD). 
Thus, DC-DGD enjoys the {\em identical} structural complexity as that of the original DGD.

2) DC-DGD is {\em memory-efficient}: In DC-DGD, each node only needs to store three local variables: $\x_{i,t},$ $\y_{i,t}$ and $\z_{i,t}.$
This is in stark contrast to some DGD-based algorithms, e.g., ADC-DGD\cite{zhang2018compressed} and DCD-PSGD\cite{tang2018communication}, where each node needs to store all values of the previous iteration from its neighbors, which is unscalable for large and dense networks where node degrees are high. 

%3) Due to SNR-constrained compressors, the noise power in the compressed local differential $C(\d_{i,t})$ is constrained by the magnitude of $\d_{i,t}$.
%Hence, if $\d_{i,t}$ is converging to zero (to be proved soon), the compression noise is also decreasing.
%Thus, {\em no} extra effort is required to tame the noise power thanks to this {\em self-compression-noise-power-reduction effect}.

3) Compared to the original DGD algorithm and many of its variants, a notable difference in DC-DGD is that the gradient $\nabla f_i(\x_{i,t})$ in Step 3-c) is calculated based on an {\em inexact} update from $\x_{i,t-1}$ and the compressed differential $C(\d_{i,t})$ (i.e., Step 3-a)), rather than using an exact update.
This is derived from the convergence of a chosen Lyapunov function (to be defined soon).
Interestingly, we will show that this modification does not harm the algorithm's convergence speed because the difference between inexact and exact updates is negligible when the Lyapunov function is near convergence.

\smallskip
Before we prove the convergence of DC-DGD, it is insightful to  offer some  intuitions on why DC-DGD retains most of the simple structural properties of the original DGD and does {\em not} need extra mechanism/parameter(s) to tame compression noises.
%For simplificity, we let $F(\x) \triangleq \sum_{i=1}^{n} f_i(x_i)$.
First, we define the following Lyapunov function:
\begin{align} \label{eqn_Lyapunov_fn}
L_{\alpha_t}(\x) & \triangleq \frac{1}{2}\x^\top (\I-\W\otimes \I_d)\x + \alpha_t f(\x). 
\end{align}
We note that $L_{\alpha_t}(\x)$ is also used for proving the convergence of several other DGD-based algorithms (e.g., \cite{yuan2016convergence,zeng2018nonconvex}).
To understand our DC-DGD algorithm, we rewrite its updates Steps 3-a) -- 3-d) in the following vector form:
\begin{align}\label{Eq: updating_DC-DGD}
\left\{
\begin{aligned}
& \x_t = \x_{t-1} + C(\d_{t}), \\
& \y_t = \y_{t-1} + (\W\otimes \I_d) C(\d_{t}), \\
& \z_{t+1} = \y_{t} - \alpha_{t} \nabla f(\x_t), \\ 
& \d_{t+1} = \z_{t+1} - \x_t,
\end{aligned}
\right.
\end{align}
where $\y\!\triangleq\! [\y_1^{\top},\ldots, \y_n^{\top}]^{\top},$
$\z\!\triangleq\! [\z_1^{\top},\ldots, \z_n^{\top}]^{\top}$
and $\d \!\triangleq\! [\d_1^{\top},\ldots, \d_n^{\top}]^{\top}$.
Note that with $\y_0 \!=\! \0,$ we have $\y_t \!=\! (\W\otimes \I_d) \x_t$ by induction.
Hence, we can rewrite the updates as:
\begin{align*} %\label{Eq: updating_DC-DGD2}
\left\{
\begin{aligned}
&\x_t = \x_{t-1} + C(\d_{t}) = \x_{t-1} + \z_{t} - \x_{t-1} + \bepi_t = \z_{t} + \bepi_t, \\
&\z_{t+1} = (\W\otimes \I_d)\x_{t} - \alpha_t \nabla F(\x_t) = \x_t - \nabla L_{\alpha_t}(\x_t), \\ 
& \d_{t+1} = \z_{t+1} - \x_t = - \nabla L_{\alpha_t}(\x_t),
\end{aligned}
\right.
\end{align*}
where $\bepi_t$ is a compression noise satisfying $\mathbb{E}[\bepi_t]=\0$ and $\mathbb{E}[\|\bepi_t\|^2] \!\le\! (1/\eta) \|\d_t\|^2 \!=\! (1/\eta) \|\nabla L_{\alpha}(\x_t)\|^2.$ 
That is, the power of the noise $\bepi_t$ depends on the difference between two successive iterations, which in turn is the gradient of the Lyapunov function $\nabla L_{\alpha_t}(\x_t)$.
As the algorithm converges (to be proved soon), 
$\nabla L_{\alpha_t}(\x_t) \rightarrow \0$ implies that $\mathbb{E}[\|\bepi_t\|^2] \rightarrow \0$.
Hence, {\em no} extra effort is required to tame the noise power thanks to this {\em self-compression-noise-power-reduction effect}.

\subsection{Main Theoretical Results}\label{sec:main_results}

In this subsection, we will establish the convergence of the proposed DC-DGD algorithm.
Our convergence results are proved under the following mild assumptions:
\begin{assumption}\label{Assumption: Objective}
The local objective functions $f_i(\cdot)$ satisfies:
\begin{list}{\labelitemi}{\leftmargin=1em \itemindent=-0.5em \itemsep=.2em}
\item (Lower boundedness) There exists an optimal $\x^*$ with $\|\x^*\|<\infty$ such that $f(\x) \ge f(\x^*)$, $\forall \x$;
\item (Lipschitz continuous gradient) there exists a constant $L > 0$ such that $\forall \x_{1},\x_{2},$ $\|\nabla f_i(\x_{1})-\nabla f_i(\x_{2})\|\le L\| \x_{1} - \x_{2} \|,~ \forall i$;
\item (Bounded gradient) there exists a constant $D > 0$ such that for all $\x$, $\|\nabla f_i(\x)\| \le D$, $\forall i$.
\end{list}
\end{assumption}

Note that the first two bullets are standard in convergence analysis: 
The first one ensures the existence of optimal solution and the second guarantees the smoothness of the local objectives. 
The third bullet is needed to bound the deviation of local copies to their mean (cf. Theorem \ref{Theorem: Theorem2}).
It is equivalent to $f_i(\cdot)$ being $D$-Lipschitz continuous.
This mild assumption has been widely adopted in analyzing non-convex optimization algorithms in the literature (see, e.g.,\cite{jiang2017collaborative,zhou2018generalization,reddi2016stochastic}). 
%Another class of widely used technique assumption is the coercivity assumption in \cite{zeng2018nonconvex} and its variant in \cite{zhang2018compressed}, which can derived a upper bound for the graidents. 

To show the convergence of DC-DGD, we will show that the iterates $\{\x_t\}_{t=1}^{\infty}$ and the gradient $\{\nabla f(\x_t)\}_{t=1}^{
\infty}$ are bounded over all iterations, and the summation of the gradients of the Lyapunov function over the iterations is also bounded.
%%%%%%%%%%%%%%%%%%%%%%%%%%%%%%%%%%%%% 
\begin{myTheo}\label{Theorem: Theorem1}
Under Assumption \ref{Assumption: Objective}, if a constant step-size $\alpha \le (\lambda_N(\eta+1)+\eta-1)/L(1+\eta)$ is used, where $\eta$ is the SNR threshold satisfying $\eta > (1-\lambda_N)/(1+\lambda_N)$, then the gradients of the Lyapunov function $L_\alpha$ is bounded, i.e., 
\begin{equation*}
\sum_{\tau=0}^{t}\mathbb{E}[\|\nabla L_{\alpha}(\x_\tau)\|^2] \le  \frac{2\alpha\big(f(\0)-f(\x^*)\big)}{ 1+ \lambda_N - \alpha L - (1-\lambda_N+\alpha L)/\eta} .
\end{equation*}
\end{myTheo}
Note that Theorem~\ref{Theorem: Theorem1} has a key condition on the SNR threshold: $\eta > (1-\lambda_N)/(1+\lambda_N)$. 
This SNR lower bound is to guarantee the feasible domain for the step-size $\alpha.$
Interestingly, it can be seen that as $\lambda_N \rightarrow 1$ (i.e., a sparse consensus matrix $\W$), the lower bound for SNR $\eta$ shrinks to zero,
meaning that as the network gets {\em sparser}, we could adopt compressors with {\em larger compression ratios}.

Next, we bound the derivation of each local copy from the mean of all local copies in any iteration $t$:
%%%%%%%%%%%%%%%%%%%%%%%%%%%%%%%%%%%%% 
\begin{myTheo}\label{Theorem: Theorem2}
Under Assumption \ref{Assumption: Objective} and with the same step-size and SNR selections as in Theorem~\ref{Theorem: Theorem1}, in each iteration $t$, the deviations of local copies from the mean can be bounded as:
\begin{align*}
\mathbb{E}[\|\x_t \!-\! \bar{\x}_t \|^2]  \!\le\! \bigg( \frac{\alpha N D}{1 \!-\! \beta} \bigg)^2 \!+\! \sum_{\tau=1}^{t}\beta^{2(t-\tau)}  \mathbb{E}[\|\nabla L_{\alpha}(\x_{\tau-1})\|^2]/\eta, 
\end{align*}
where $\bar{\x}_t = \1\1^\top\x_t/N$ and $\beta = \max\{|\lambda_2|,|\lambda_N|\}.$
\end{myTheo}
Theorem \ref{Theorem: Theorem2} requires that $\mathbb{E}[\nabla L_{\alpha}(\x_t)]$ is bounded, which is guaranteed by Theorem \ref{Theorem: Theorem1}.
Lastly, based on Theorems~\ref{Theorem: Theorem1} and \ref{Theorem: Theorem2}, we show that DC-DGD converges to an error ball of the global objective's stationary point at rate $O(1/t)$:
%%%%%%%%%%%%%%%%%%%%%%%%%%%%%%%%%%%%% 
\begin{myTheo}\label{Theorem: Theorem3}
Under Assumption \ref{Assumption: Objective}, if the step-size satisfies $\alpha \le (\lambda_N(\eta+1)+\eta-1)/L(1+\eta),$ then it holds that 
\begin{equation*}
\sum_{\tau=0}^{t}\mathbb{E}[\|\nabla f(\bar{\x}_{\tau})\|^2] 
\le C_1(\alpha,\beta)[f(\0) -f(\x^*)] + \frac{\alpha^2N^2D^2L}{(1-\beta)^2}t,
\end{equation*}
where $C_1(\alpha,\beta)=4(\frac{\alpha}{(1-\beta^2)}+\frac{L}{2})/[(1+ \lambda_N - \alpha L)\eta - (1-\lambda_N+\alpha L)]+\frac{2N}{\alpha}.$ 
Thus, DC-DGD converges at rate $O(1/t)$ to an error ball that depends on parameters $(\alpha,N,D,L,\beta)$:
\begin{equation*}
\min_{\tau=0,\cdots,t} \!\!\!\! \mathbb{E}[\|\nabla f(\bar{\x}_{\tau})\|^2] 
\!\le\! \frac{C_1(\alpha,\beta)[f(\0) \!-\! f(\x^*)]}{t} + \frac{\alpha^2N^2D^2L }{(1 \!-\! \beta)^2}.
\end{equation*}
\end{myTheo}
Note that in Theorem \ref{Theorem: Theorem3}, similar to the original DGD algorithm, the size of the error ball is determined by two terms: The first one is a convergence error with sublinear diminishing rate $O(1/t)$; The second term is the approximation error affected by the step-size and the network structure (characterized by $N$ and $\beta$). 
Therefore, to reach an optimal solution, the step-size $\alpha$ needs to be small so that the second term is close to zero. 
However, as $\alpha \rightarrow 0,$ the coefficient for the convergence error $C_1(\alpha,\beta) \approx 2/\alpha \rightarrow \infty,$ which in turn requires more iterations for shrinking the first term.
 
The next result shows that with diminishing step-size $\alpha_t=O(1/t^{1/3}),$ DC-DGD converges to a first-order stationary point (optimal solution in convex problems) at rate $O(1/t^{2/3})$:
%%%%%%%%%%%%%%%%%%%%%%%%%%%%%%%%%%%%%
\begin{myCor}\label{Corollary: Corollary1}
Let $\alpha_t = (C_2/t)^{1/3}$, where $C_2 \triangleq (f(\0) - f(\x^*))(1-\beta)^2/D^2N^2L,$ and $\alpha_t \le (\lambda_N (\eta+1)+1-\eta)/L(1+\eta)$, then the convergence rate of DC-DGD is: 
\begin{equation*}
\!\! \min_{\tau\in[0,t]} \!\!\! \mathbb{E}[\|\nabla f(\bar{\x}_{\tau})\|^2] 
\!\!\le\!\! \frac{3 \big({f}(\0) \!\!-\!\! {f}(\x^*)\big)^{2/3} (D^2 N^2 L)^{1/3}}{(1-\beta)^{2/3}t^{2/3}}+\!O\Big( \frac{1}{t} \Big).
\end{equation*}
\end{myCor}

%The proofs of the main results could be found in the appendix.

\subsection{Proofs of the Main Theoretical Results} \label{sec:proof_sketches}

Due to space limitation, we provide proof sketches of the main theoretical results in this subsection.

\begin{proof}[Proof Sketch of Theorem~\ref{Theorem: Theorem1}]
Let $\mathcal{F}_t \!\triangleq\! \sigma (\x_1,\!\cdots,\! \x_t)$ denote a filtration.
It can be shown that the Lyapunov function $L_{\alpha}(\x)$ has $(1\!-\!\lambda_n\!+\!\alpha L)$-Lipschitz gradients.
It then follows that:
\begin{multline*}
L_\alpha(\x_{t+1}) \le L_\alpha(\x_{t}) - \langle \nabla L_{\alpha}(\x_{t}), \nabla L_{\alpha}(\x_{t})-\bepi_{t+1} \rangle + \\
\frac{(1-\lambda_N+\alpha L)}{2} [\|\nabla L_{\alpha}(\x_{t})\|^2+\|\bepi_{t+1}\|^2-2\langle \nabla L_{\alpha}(\x_{t}), \bepi_{t+1} \rangle].
\end{multline*}
Taking conditional expectation and using the properties of SNR-constrained unbiased compressors yield:
$\mathbb{E}[L_\alpha(\x_{t+1})|\mathscr{F}_t] \leq L_\alpha(\x_{t}) + \frac{1}{2}[(\alpha L -\lambda_N - 1)
+ \frac{(1-\lambda_N+\alpha L)}{\eta}]\|\nabla L_{\alpha}(\x_{t})\|^2$. 
Since $\eta > (1-\lambda_N)/(1+\lambda_N),$ we have $(\lambda_N(\eta+1)+\eta-1)/L(1+\eta)>0.$ 
Then, by setting step-size $\alpha$ as stated in the theorem, we have $[\alpha L -\lambda_N \!-\! 1 \!+\! (1 \!-\! \lambda_N \!+\! \alpha L)/\eta] < 0$. 
It then follows that
$-[\alpha L -\lambda_N - 1 + (1-\lambda_N +\alpha L)/\eta]\|\nabla L_{\alpha}(\x_{t})\|^2 \le 2(L_\alpha(\x_{t})  -\mathbb{E}[L_\alpha(\x_{t+1})|\mathscr{F}_t])$.
Taking full expectation on both sides and telescoping from $0$ to $t$, we have:
\begin{multline}
-[\alpha L -\lambda_N - 1 + (1-\lambda_N+\alpha L)/\eta] \times \\
\sum_{\tau=1}^{t}\mathbb{E}[\|\nabla L_{\alpha}(\x_{t})\|^2]  \le 2(L_\alpha(\x_0)  -\mathbb{E}[ L_\alpha(\x_{t+1})]).
\end{multline}
Since $\mathbb{E}[ L_\alpha(\x_{t+1})] \ge \alpha\mathbb{E}[ f(\x_{t+1})] \ge \alpha \sum_{i=1}^{n}f_i(\x^*)$, after rearranging terms, we can conclude that:
\begin{align*}
\sum_{\tau=1}^{t}\mathbb{E}[\|\nabla L_{\alpha}(\x_{t})\|^2]
\leq \frac{2\alpha( \sum_{i=1}^{N} f_i(\0) - \sum_{i=1}^{N} f_i(\x_*))}{ 1+ \lambda_N - \alpha L
- (1-\lambda_N+\alpha L)/\eta},
\end{align*}
and the proof is complete.
\end{proof}

\begin{proof}[Proof Sketch of Theorem~\ref{Theorem: Theorem2}]
For notation convenience, We let $\tilde{\W} \triangleq \W \otimes \I_d.$
From (\ref{Eq: updating_DC-DGD}), we can obtain:
\begin{align}
\left\{
\begin{aligned}
\x_1 &= \tilde{\W}\x_0 - \alpha_0 \nabla f(\x_0)-\bepi_1= - \alpha_0 \nabla f(\x_0)-\bepi_1, \\
\x_2 &= \tilde{\W}\x_1 - \alpha_1 \nabla f(\x_1)-\bepi_2\notag\\
&= -\tilde{\W} \alpha_0\nabla f(\x_0) - \alpha_1\nabla f(\x_1)-\tilde{\W}\bepi_1-\tilde{\W}\bepi_2,\\
& \hspace{.08in} \vdots\\
\x_t & = -\sum_{\tau=0}^{t-1}\alpha \tilde{\W}^{t-\tau-1}\nabla f(\x_\tau)-\sum_{\tau=1}^{t}\tilde{\W}^{t-\tau}\bepi_\tau.
\end{aligned}
\right.
\end{align}
Using the above equations, we can derive the following inequality for the deviation from the mean $\bar{\x}_t$:
\begin{align*}
&
\|\x_t - \bar{\x}_t \|^2 =  \|\x_t - (1/N) \1\1^\top\x_t\|^2 \notag \\
&\leq 2 \big\| \sum\nolimits_{\tau=0}^{t-1}\alpha (\tilde{\W}^{t-\tau-1}-(1/N)\1\1^\top)\nabla f(\x_\tau) \big\|^2  \notag \\
&
+
2\sum_{\tau=1}^{t}\|(\tilde{\W}^{t-\tau}-(1/N)\1\1^\top)\bepi_\tau\|^2 \notag\\
&
+2\sum_{\tau=1}^{t}\sum_{s=1,s \neq \tau}^{t} \!\!\! \bigg\langle \bigg(\tilde{\W}^{t-\tau}-\frac{\1\1^\top}{N}\bigg) \bepi_\tau, \bigg(\tilde{\W}^{t-s}-\frac{\1\1^\top}{N}\bigg)\bepi_s \bigg\rangle.
\end{align*}
Taking the expectation on both sides, noting $\mathbb{E}[\bepi_t] = \0$, and after some algebraic manipulations, we arrive at:
\begin{align*}
\mathbb{E}[\|\x_t \!-\! \bar{\x}_t \|^2]  \!\le\! \bigg( \frac{\alpha N D}{1 \!-\! \beta} \bigg)^2 \!+\! \sum_{\tau=1}^{t}\beta^{2(t-\tau)}  \mathbb{E}[\|\nabla L_{\alpha}(\x_{\tau-1})\|^2]/\eta, 
\end{align*}
which completes the proof.
\end{proof}

\begin{proof}[Proof Sketch of Theorem \ref{Theorem: Theorem3}]
First, we prove a key descending inequality on $\bar{\x}_t=\frac{1}{N}\sum_{i=1}^{N}{\x}_{i,t}$.
From the update rule $\x_{t+1}=\tilde{\W}\x_{t}-\alpha \nabla f({\x_t})-\bepi_{t+1}$, we have $\bar{{\x}}_{t+1} = \bar{\x}_{t} - \frac{\alpha}{N}\sum_{i=1}^{N} \nabla f_i(\x_{i,t}) - \bar{\bepi}_{t+1}$.
It then follows that:
\begin{align*}
&\barf(\bar{{\x}}_{t+1}) \le \barf(\bar{{\x}}_{t}) - \langle \nabla \barf(\bar{{\x}}_{t}) ,  \frac{\alpha}{N}\sum_{i=1}^{N}\nabla f_i({\x}_{i,t}) + \bar{\bepi}_{t+1}\rangle \notag\\
&
+ \frac{L}{2} \bigg[\Big\|\frac{\alpha}{N} \nabla f({\x}_{i,t})\Big\|^2 + \|\bar{\bepi}_{t+1}\|^2 + 2 \Big\langle \frac{\alpha}{N} \nabla f({\x}_{i,t}) , \bar{\bepi}_{t+1} \Big\rangle \bigg].
\end{align*}
where $\barf(\x) = \frac{1}{N}\sum_{i=1}^{N} f_i(x_i).$
Taking the conditional expectation on both sides and after some algebraic manipulations, we can show that
\begin{multline*}
\mathbb{E}[\barf(\bar{{\x}}_{t+1})|\mathscr{F}_t]
\le 
\barf(\bar{{x}}_{t}) - \frac{\alpha}{2}\|\nabla \barf(\bar{{x}}_{t})\|^2 + \notag\\
\frac{\alpha}{2} \|\frac{1}{n}\sum_{i=1}^{n}\nabla f_i({x}_{i,t})-\nabla \barf(\bar{{x}}_{t})\|^2 + \frac{L}{2n^2\eta}\|\nabla L_\alpha(\x_t)\|^2.
\end{multline*}
Taking the full expectation, telescoping the inequality from $\tau=0$ to $t$, and after further algebraic manipulations, we have:
\begin{multline*}
\frac{\alpha}{2}\sum_{\tau=0}^{t}\mathbb{E} [\|\nabla f(\bar{{\x}}_{\tau})\|^2] 
\le \bigg[\frac{1}{N} + \bigg(\frac{\alpha L}{(1-\beta^2)N^2\eta}+\frac{L}{2N^2\eta}\bigg) \times \\
\frac{2\alpha}{1+ \lambda_N - \alpha L - (1-\lambda_N + \alpha L)/\eta}\bigg] [f(\0) - f(\x_*)]+ \frac{\alpha^3D^2Lt}{(1-\beta)^2},
\end{multline*}
which, after further rearrangements, yields the result stated in the theorem.
This completes the proof.
\end{proof}

\section{A Hybrid Compression Design under DC-DGD for Communication Cost Minimization}\label{Section: Hybrid}

Inspired by previous theoretical insights, in this section, our goal is to design a {\em hybrid} SNR-constrained compression scheme to achieve high communication cost saving, while having a controllable SNR.
Recall from Section~\ref{sec:formulation} that the sparsifier can control the compression noise power by adjusting the probability $p$ and the expected communication cost for a $d$-dimensional vector is $d[c_1p +c_0(1-p)],$ where $c_1$ is the cost for sending a floating number and $c_0$ is the cost for value $0$. 
Therefore, if the SNR $\eta$ threshold is large, the communication cost will be close to sending uncompressed copy $dc_1$.
For the ternary operator, its compression noise power is $\sum_{i=1}^{d} |z_i|(\|{\z}\|_\infty-|z_i|),$ which is {\em not} directly controllable by any parameter.
The communication cost is $c_1+(d-1)c_0',$ where $c_0'$ is the cost for the ternary values $\{-1,0,1\}$. 

In general, the communication cost of a ternary-compressed vector is much smaller than that of the sparse-compressed vector: 
For example, if using $32$-bit floating numbers and one bit for the zero value, the cost for a $d$-dimensional sparse compressed vector is $[32p+(1-p)]d$. In contrast, for the ternary operator, the cost will be $32+2(d-1)$ if using $32$-bit floating numbers and two bits for the ternary values. 
With a larger SNR threshold $\eta$ (i.e., larger $p$) and high dimensionality $d$, the communication cost of the ternary compressor is much smaller.
Therefore, to have a {\em controllable} compression noise power as well as high communication cost savings, a promising solution is to {\em combine} the sparse and the ternary compressors.

To this end, consider a $d$-dimensional vector $\z = [z_1,\cdots,z_d]^\top$.
We can sort and rearrange the elements of $\z$ in descending order of magnitude to have: $z_{[1]},\ldots,z_{[d]}$, with $|z_{[i]}| \ge |z_{[i+1]}|,$ $i = 1,\ldots,d-1$.
For the first largest $s_1$ elements, we apply the ternary compressor, while for the rest of the elements, we use the sparse compressor, i.e.,
\begin{align}
&\hspace{-.005in}\underbrace{z_{[1]}, ~z_{[2]},~ \cdots, ~z_{[s_1-1]}, ~z_{[s_1]}}_{\text{ternary compression}},\underbrace{z_{[s_1+1]},~\cdots,~z_{[d-1]}, ~z_{[d]}}_{\text{sparsifier compression}} \Rightarrow \notag\\
&\hspace{-.005in}\underbrace{z_{[1]}, ~~~0,~ ~~\cdots, ~ ~~~-1, ~~~~~1}_{\text{ternary compressed}}, \underbrace{\frac{z_{[s_1+1]}}{p},~\cdots, ~~~~~~0 , ~~\frac{z_{[d]}}{p}}_{\text{sparsifer compressed}}\nonumber
\end{align}
As a result, the compression noise power levels of the first $s_{1}$ largest elements and the rest are $\sum_{i=1}^{s_1} |{z}_{[i]}|(|{z}_{[1]}|-|z_{[i]}|)$ and $(1/p-1)\sum_{i=s_1+1}^{d}z_{[i]}^2$, respectively. 
In order to ensure the effective SNR of the hybrid scheme satisfies $\eta > C$ for some lower bound $C$, we have:
\begin{align}
(\text{ternary})&:~|{z}_{[i]}|(|{z}_{[1]}|-|z_{[i]}|) < (1/C) z_{[i]}^2,~ \forall i \le s_1 \label{Eq: ternary}\\
(\text{sparsifer})&:~ (1/p-1)z_{[i]}^2 < (1/C) z_{[i]}^2, ~\forall i > s_1 \label{Eq:sparse}.
\end{align} 
To satisfy (\ref{Eq: ternary}) and (\ref{Eq:sparse}), we have $s_1 = \arg\min_{i}\{|z_{[i]}| > \frac{1}{1+1/C}|z_{[1]}|\}$ and $p > \frac{1}{1+1/C}$, respectively. 
Then, on average, the compressed vector has $1+(d-s_1)p$ floating numbers and $(s_1-1)+(d-s_1)(1-p)$ ternary values, which is more efficient compared to that under the sparsifier compressor.

In fact, the hybrid compression idea above can be generalized to achieve further communication cost savings:
Instead of just using $z_{[1]}$ for the ternary compression, we can select multiple ``{\em anchor elements}'' $\{z_{[q_1]},\cdots,z_{[q_k]}\}$.
There are $s_i$ elements between $z_{[q_i]}$ and $z_{[q_{i+1}]}$.
Thus, a $d$-dimensional vector can be partitioned into $(k+1)$ groups. 
For the elements with indices in $(q_i, q_i+s_i)$, we apply the ternary compressor based on $z_{[q_i]}$.
For the remaining elements, we apply the sparsifier operator. 
Similar to (\ref{Eq: ternary}), we have
\begin{align}\label{Eq: tenary criteria}
|{z}_{[j]}|(|{z}_{[q_i]}|-|z_{[j]}|) < (1/C) z_{[j]}^2,~ \forall j \in (q_i,q_i+s_i).
\end{align}
Then, the compressed vector has $k+(d-\sum_{i=1}^{k}s_i)p$ floating and $(\sum_{i=1}^{k}s_i-k)+(d-\sum_{i=1}^{k}s_i)(1-p)$ ternary values. 
Moreover, we need to save the indices of the anchor elements,
for which we need $\lceil \log(k+1)\rceil$ bits per element.

%In the above, we give a family of the hybrid compression operators, which include the sparse operator and the ternary operator as the special cases. 
Given a SNR threshold $\eta,$ the communication saving of our hybrid compression scheme is highly dependent on the group number $k$ and the positions of the anchor elements,
which can be optimized by solving an integer programming problem. 
Take 32-bit floating numbers and 2-bit ternary values as an example. 
To achieve the maximum communication saving, the group number $k$ and the locations of the anchor elements can be determined by solving:
\begin{multline}\label{Eq: selection_obj}
\min_{k,\{z_{[q_i]}\}} \Bigg\{ 32 \underbrace{\bigg[k+\bigg(d-\sum_{i=1}^{k}s_i \bigg)p\bigg]}_{\text{Number of floating values}} + [2 + \underbrace{\lceil \log(k+1) \rceil}_{\substack{\text{cost of storing} \\ \text{anchor indices}}}] \times \\
 \underbrace{\bigg[\bigg(\sum_{i=1}^{k}s_i-k\bigg)+\bigg(d-\sum_{i=1}^{k}s_i\bigg)(1-p)\bigg]}_{\text{Number of ternary values}} \Bigg\}.
\end{multline}
Problem (\ref{Eq: selection_obj}) is an integer optimization problem, which can be shown to be equivalent to bin packing problems, thus being NP-hard. 
However, an efficient greedy heuristic algorithm can be developed by leveraging the special problem structure.
Specifically, we note that the objective function is increasing and decreasing with respect to $k$ and $\sum_{i=1}^{k}s_i$, respectively. 
Therefore, we can find anchor points $\{\z_{[q_i]}\}_{i=1}^{k}$ and their corresponding ternary sets (of size $s_i$) by checking (\ref{Eq: tenary criteria}); 
if the ternary cost of the $s_i$ elements is smaller than the sparsifier cost, we remove these $s_i$ elements from the current vector; otherwise, we use the sparsifier compressor on the current vector. 
We summarize the greedy algorithm as follows: 
\medskip
\hrule 
\vspace{.03in}
\noindent {\textbf{ Algorithm~2:}} A greedy algorithm for solving Problem~(\ref{Eq: selection_obj}).
\vspace{.03in}
\hrule
\vspace{0.1in}
\noindent {\textbf{ Initialization:}}
\begin{enumerate} [topsep=1pt, itemsep=-.1ex, leftmargin=.2in]
\item[1.] Sort and rearrange the elements of vector $\z$ in descending order of magnitude. 
\item[2.] Let $i=1$. Set the ternary set $\mathcal{T}$ as empty. 
\end{enumerate}

\noindent {\textbf{ Main Loop:}}
\begin{enumerate} [topsep=1pt, itemsep=-.1ex, leftmargin=.2in]
\item[3.] \textbf{ Inner Loop:}
\begin{enumerate} [topsep=1pt, itemsep=-.1ex, leftmargin=.4in]
\item [3.1) ] For each element $\z_{[j]}$, $j \notin \mathcal{T},$ find the set: $\mathcal{S}_j=\{{z}_{[k]}: |{z}_{[k]}|(|{z}_{[j]}|-|z_{[k]}|) < z_{[k]}^2/C, k \notin \mathcal{T}\}$.
\item [3.2) ] Set $q_i = \arg \max |\mathcal{S}_j| $ and $s_i = \max|\mathcal{S}_j|$. % with $|\cdot|$ as the cardinality.
\end{enumerate}

\item[4.] Compare the ternary cost $32+2(s_i-1)$ with the sparsifier cost $[32p+2(1-p)]s_i;$ 
\item[5.] If the ternary cost is smaller, then remove the corresponding elements from the current vector and add them to $\mathcal{T}$, let $i\leftarrow i+1$ and go to Step 3; otherwise, break the loop.
\end{enumerate}

\noindent {\textbf{Final Step:}} 
\begin{enumerate} [topsep=1pt, itemsep=-.1ex, leftmargin=.2in]
\item[6.] Apply the ternary operator to each group in $\mathcal{T}$ and the sparse operator to $\mathcal{T}^c.$
\end{enumerate}
\smallskip
\hrule
\medskip

Now, we analyze the running time complexity of the greedy algorithm.
First of all, the sorting requires $O(d\log(d))$ time. 
The worst-case number of iterations in the main loop is $O(d);$ while in each inner loop, it takes $O(d)$ steps to find the ternary set for each element. 
Hence, the overall time-complexity of Algorithm~2 is $O(d^2 + d\log(d))$.

\section{Numerical Results}\label{Section: Numerical}

In this section, we perform extensive numerical experiments to validate the performances of our proposed DC-DGD algorithm and the hybrid compression scheme.

\begin{figure*}[!ht]
%\vspace{-.1in}
    \centering
    \subfigure[]{
        \includegraphics[width=0.23\textwidth]{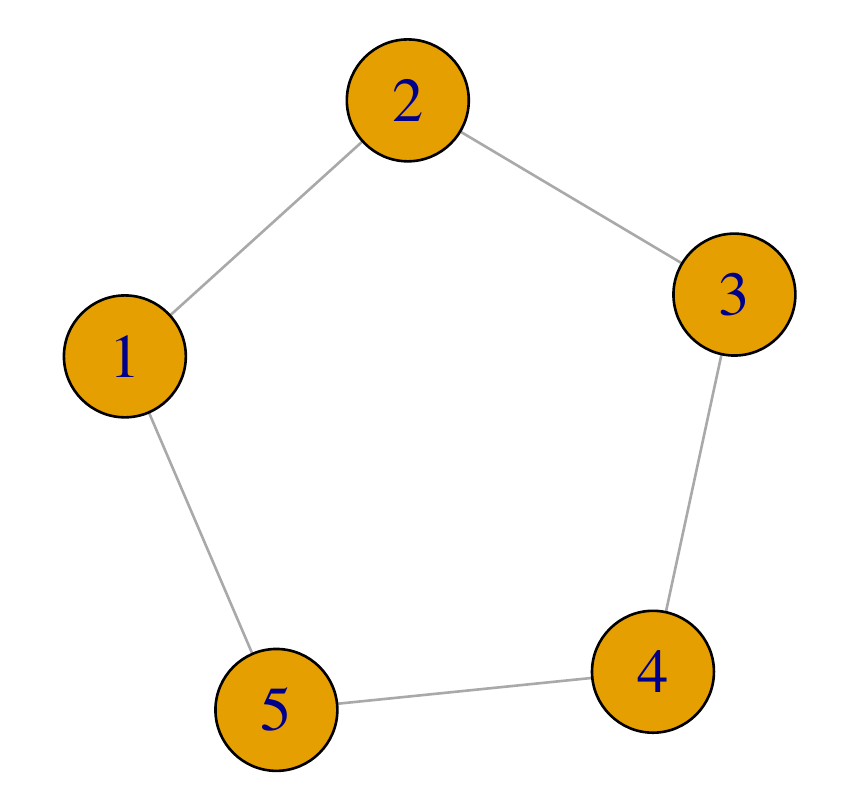}
        \label{Fig:simu1_network}
    }
    \hspace{.01\textwidth}
    \subfigure[]{
        \includegraphics[width=0.3\textwidth]{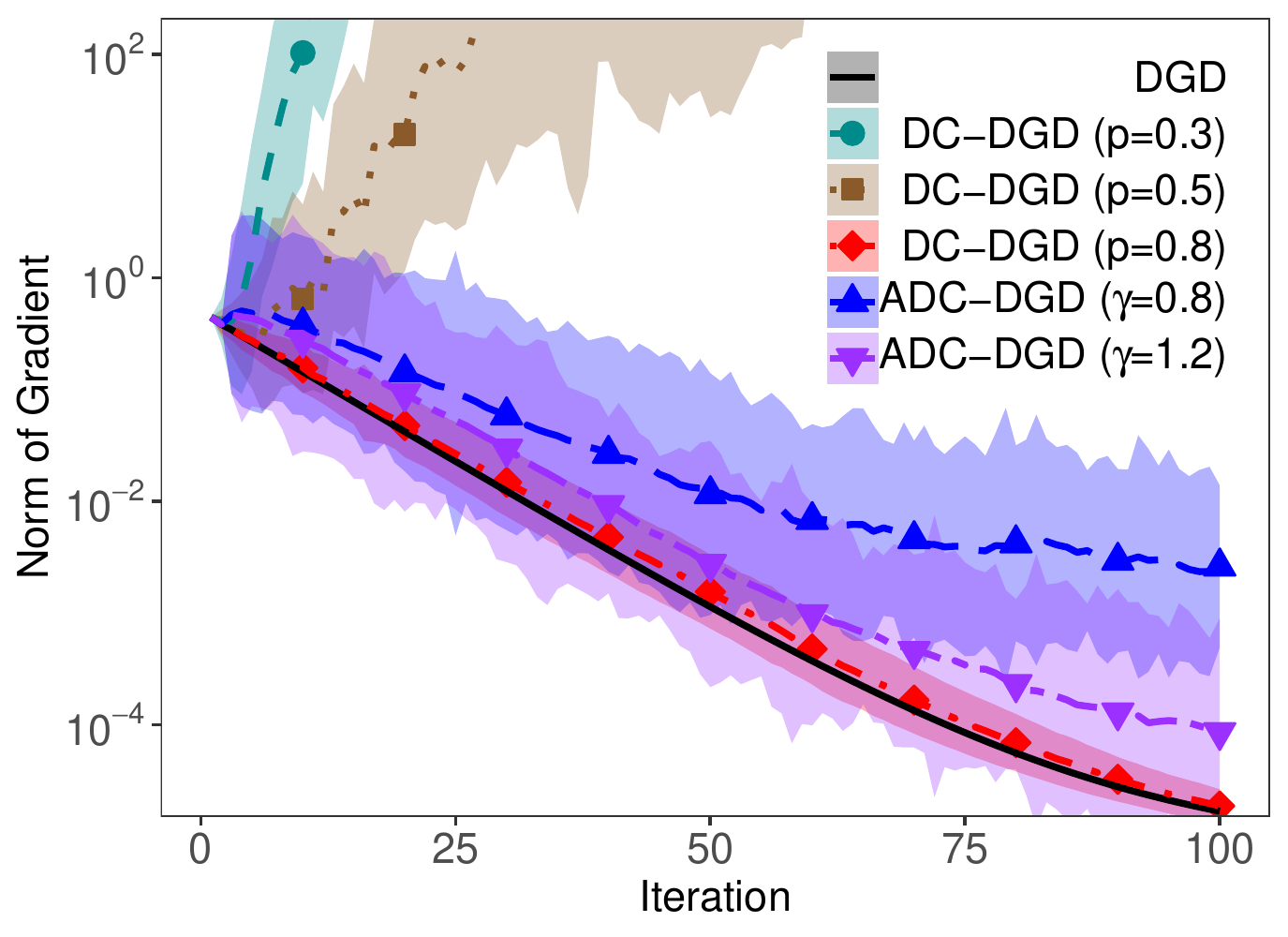}
        \label{Fig:simu1_1}
    }
    \hspace{.01\textwidth}
    \subfigure[]{
        \includegraphics[width=0.3\textwidth]{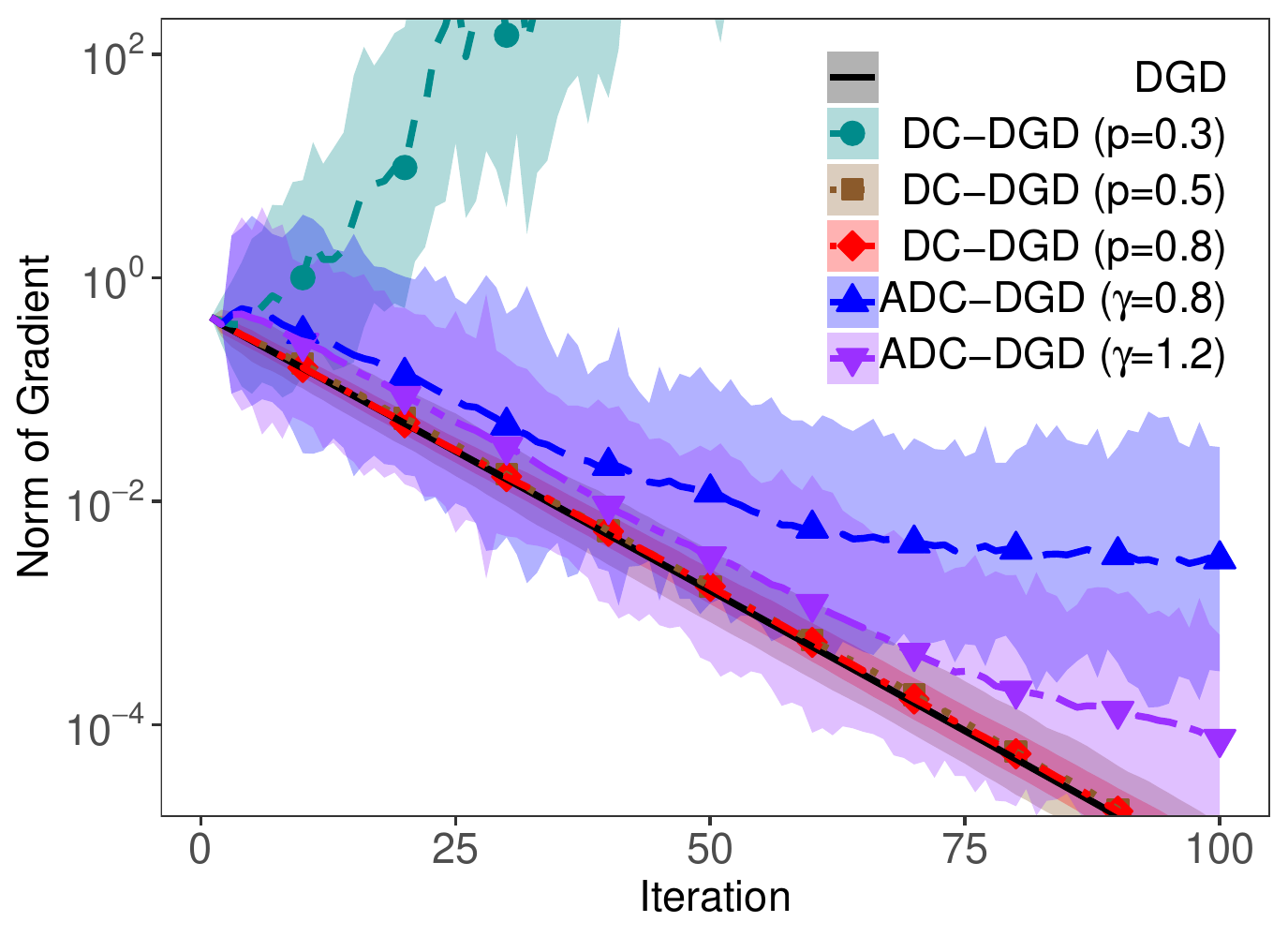}
        \label{Fig:simu1_2}
    }
%\vspace{-.1in}
    \caption{(a) The five-node circle network; (b-c) Performance comparsion: Convergence error vs Iteration with the consensus matrices $\W_1$ and $\W_2$, respectively. The black solid curve is the original DGD algorithm. The other curves represent the error averaged over 50 trials and the shaded regions indicate the standard deviations of results over random trials.}\label{Fig: Simu_1}
\vspace{-.1in}
\end{figure*}

\begin{figure*}[!ht]
%\vspace{-.1in}
     \subfigure[{\footnotesize{Case1: Bias.}}]{
        \includegraphics[width=0.3\textwidth]{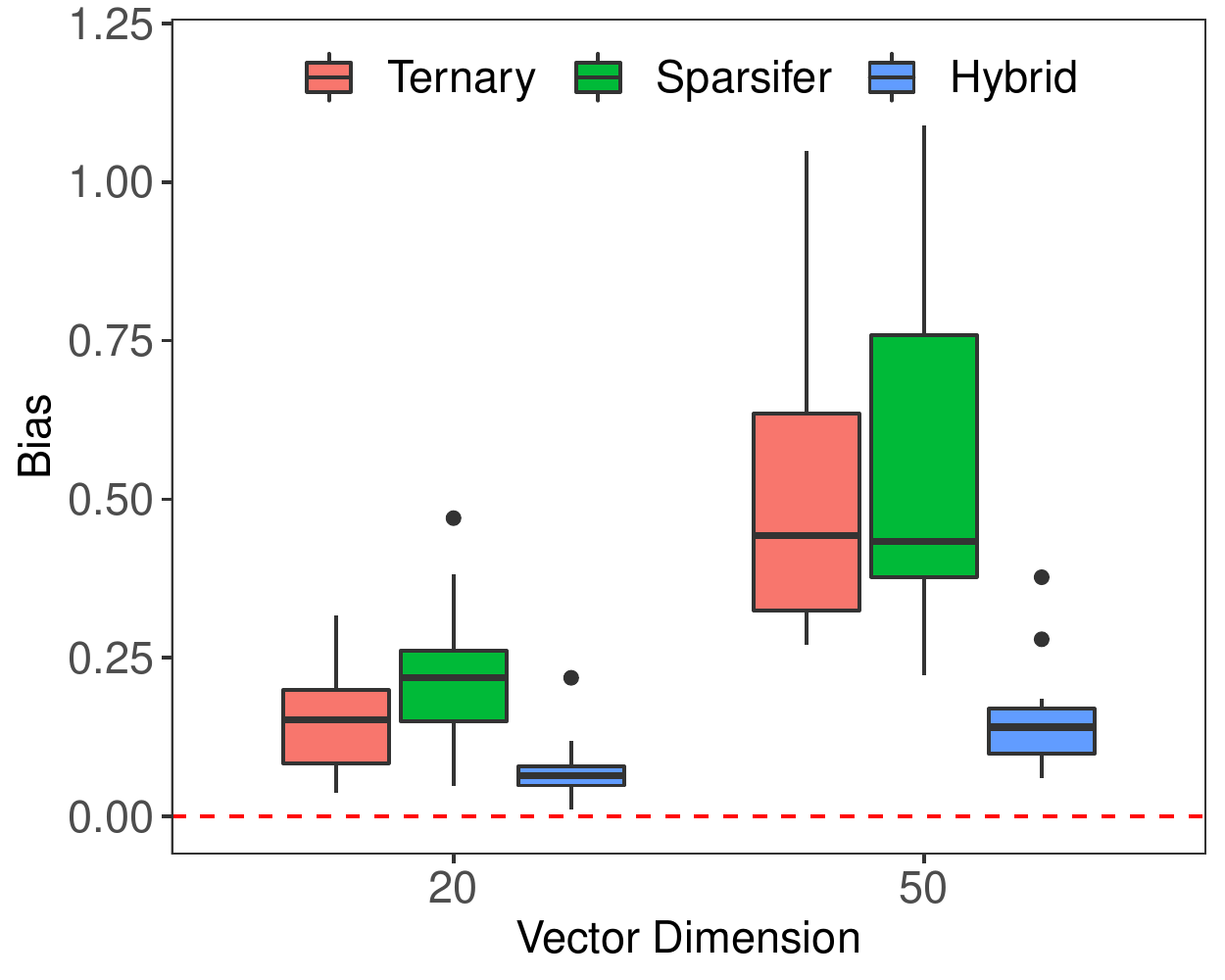}
        \label{Fig:simu2_4}
    }
    \hspace{.01\textwidth}
    \subfigure[{\footnotesize{Case1: Signal-to-Noise Ratio.}}]{
        \includegraphics[width=0.3\textwidth]{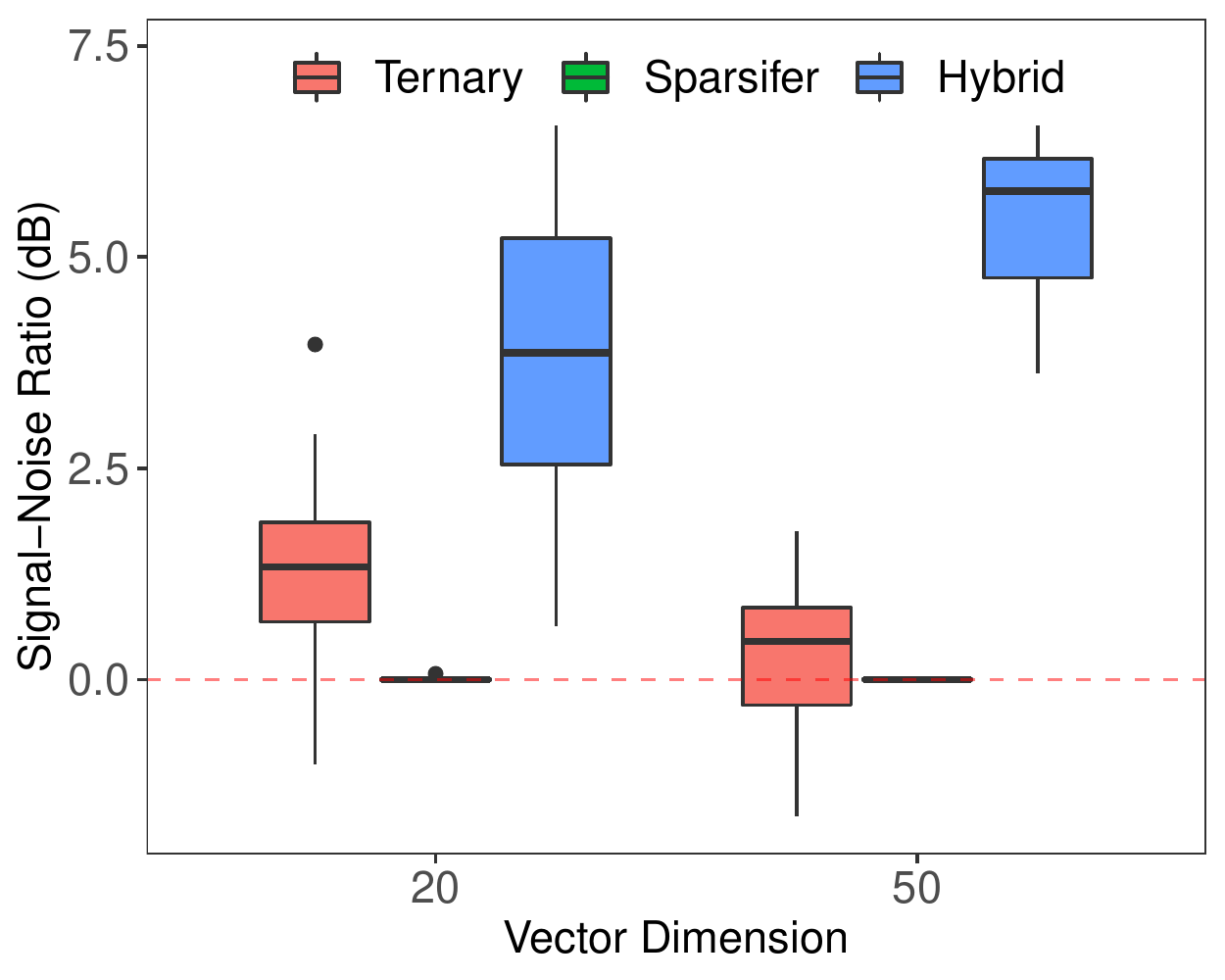}
        \label{Fig:simu2_5}
    }
    \hspace{.01\textwidth}
    \subfigure[{\footnotesize{Case1: Communication Cost.}}]{
        \includegraphics[width=0.3\textwidth]{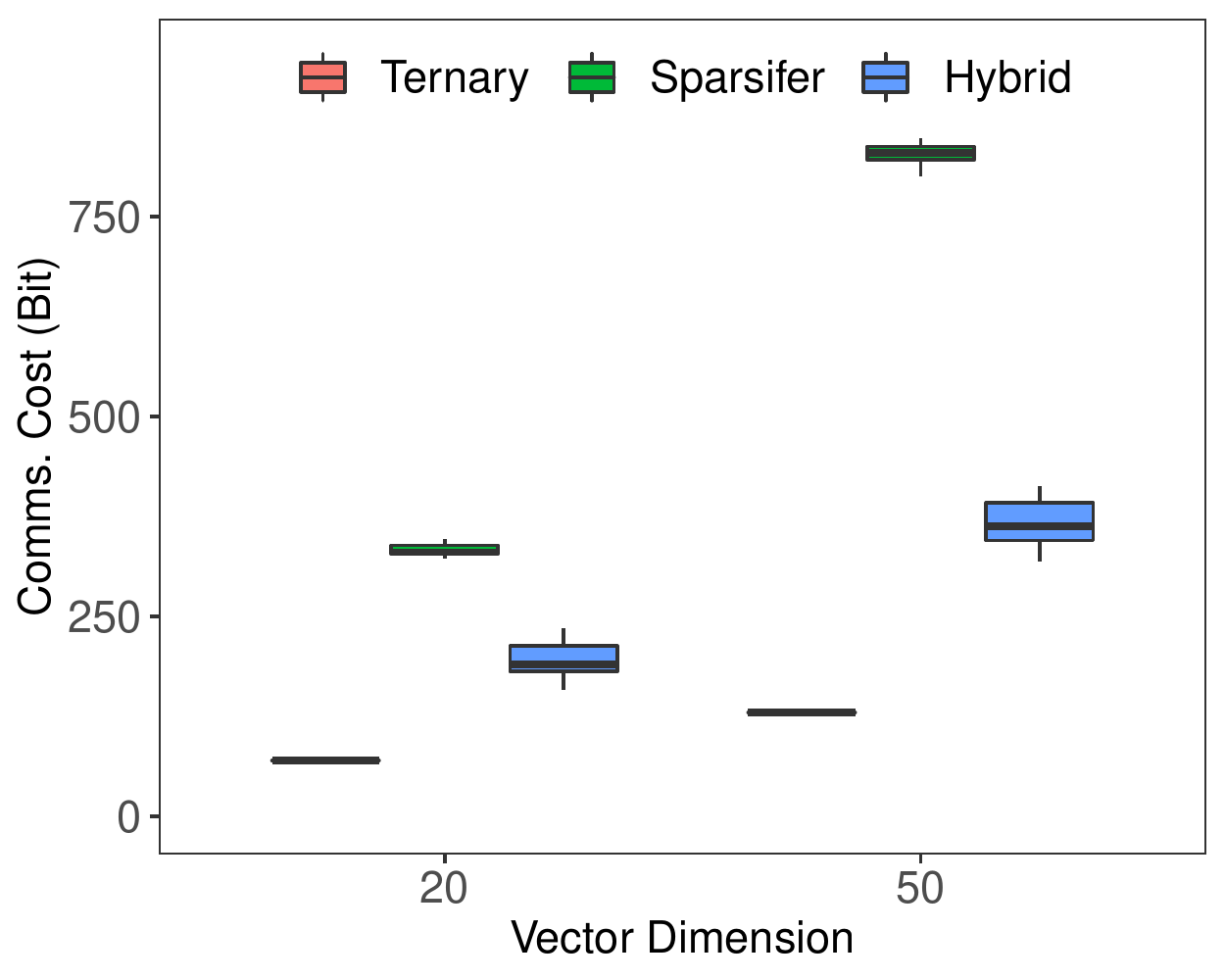}
        \label{Fig:simu2_6}
    }
        \centering
    \subfigure[{\footnotesize{Case2: Bias.}}]{
        \includegraphics[width=0.3\textwidth]{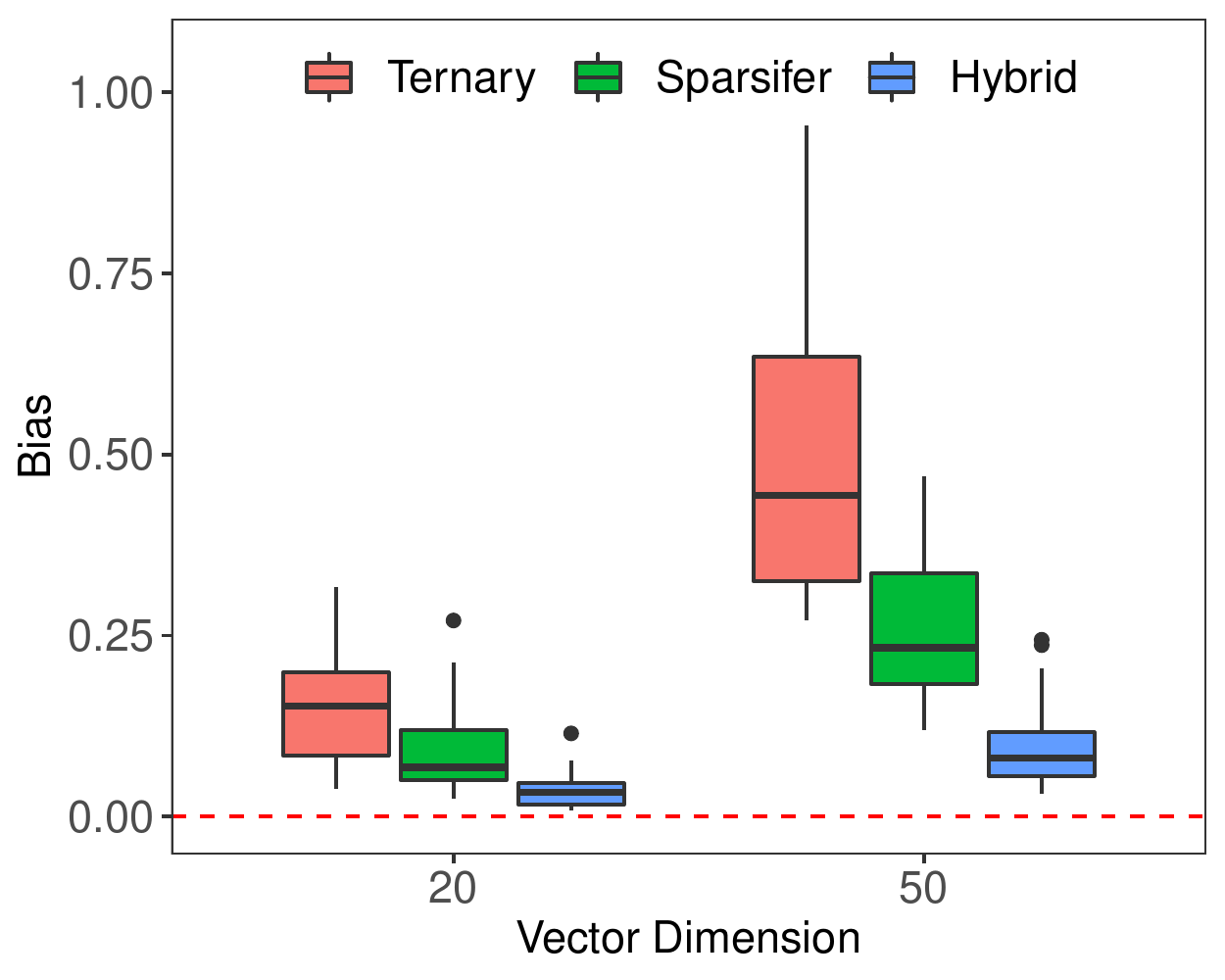}
        \label{Fig:simu2_1}
    }
    \hspace{.01\textwidth}
    \subfigure[{\footnotesize{Case2: Signal-to-Noise Ratio.}}]{
        \includegraphics[width=0.3\textwidth]{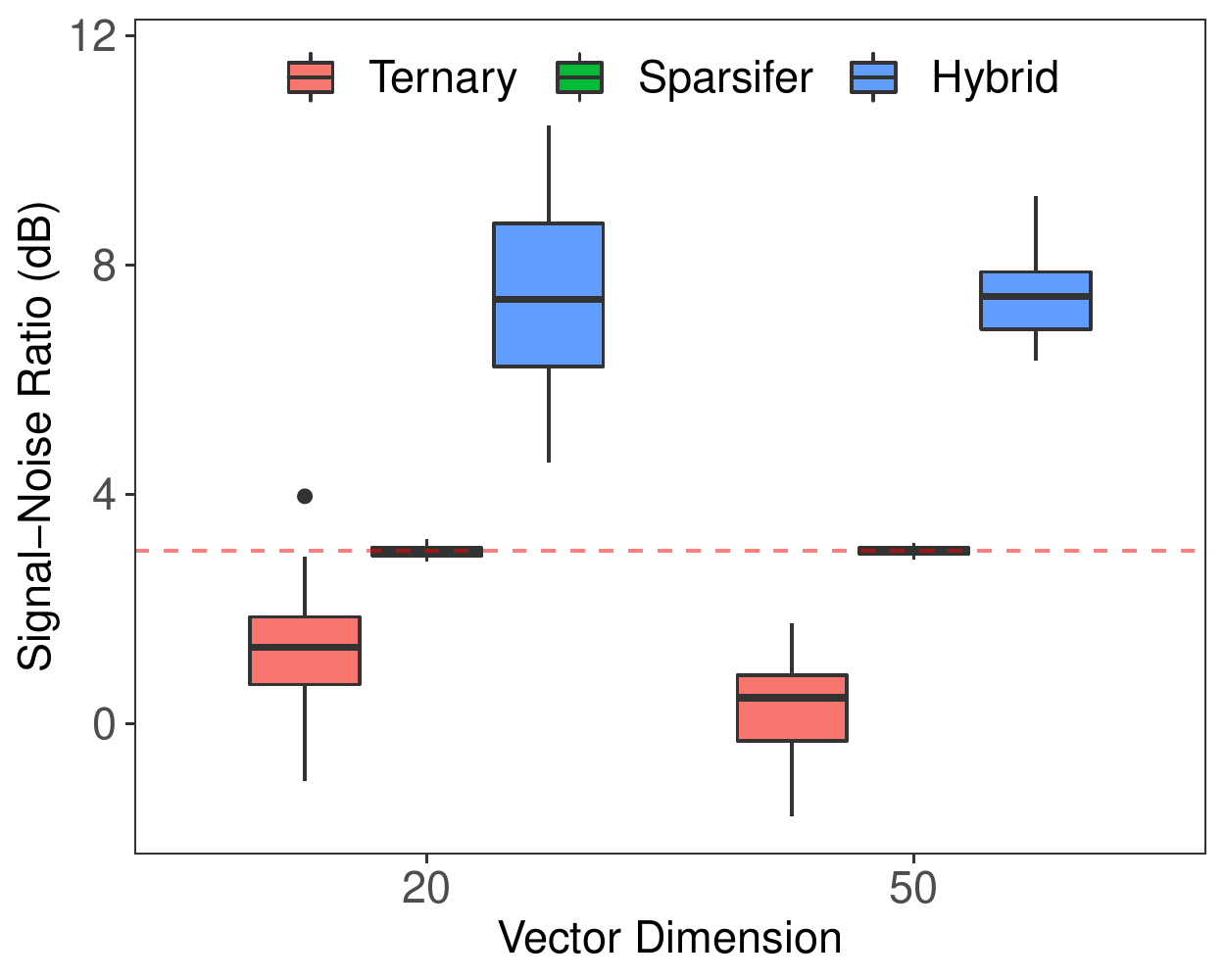}
        \label{Fig:simu2_2}
    }
    \hspace{.01\textwidth}
    \subfigure[{\footnotesize{Case2: Communication Cost.}}]{
        \includegraphics[width=0.3\textwidth]{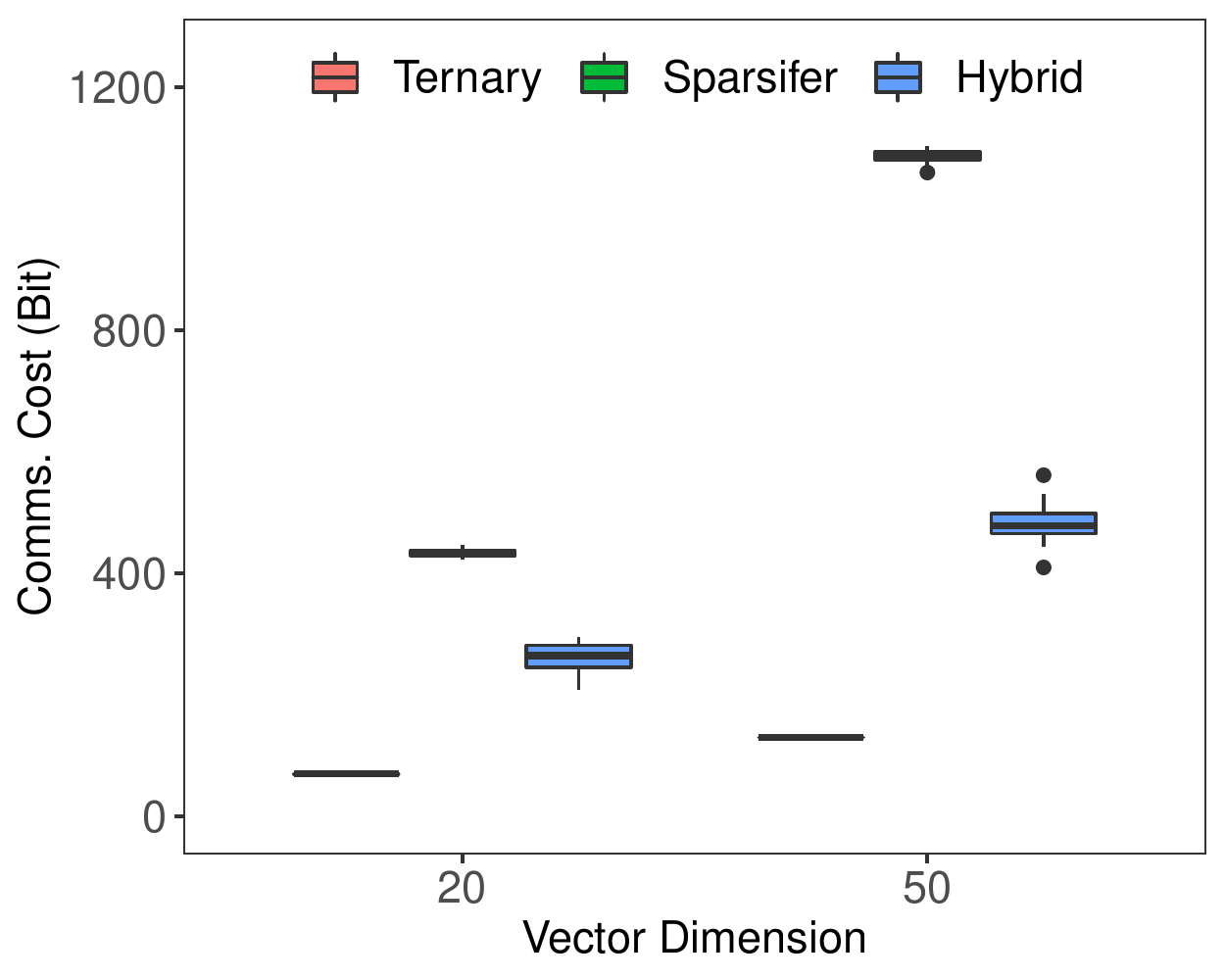}
        \label{Fig:simu2_3}
    }\\
     %\vspace{-.1in}
    \caption{Comparisons between three compressors: (a)-(c) are the boxplots for the SNR lower bound as $0$dB; and (d)-(e) are the boxplots for the SNR lower bound $3$ dB. The red dashed lines in (a) and (d) represents $0$; the red dashed lines in (b)\&(e) present the SNR lower bound $0$ dB and $3$ dB, respectively.}\label{Fig: Simu_2}
    \vspace{-.2in}
\end{figure*}

\begin{figure*}[t!]
%\vspace{-.1in}
    \centering
    \subfigure[{\footnotesize{Network Topology of Case1.}}]{
        \includegraphics[width=0.22\textwidth]{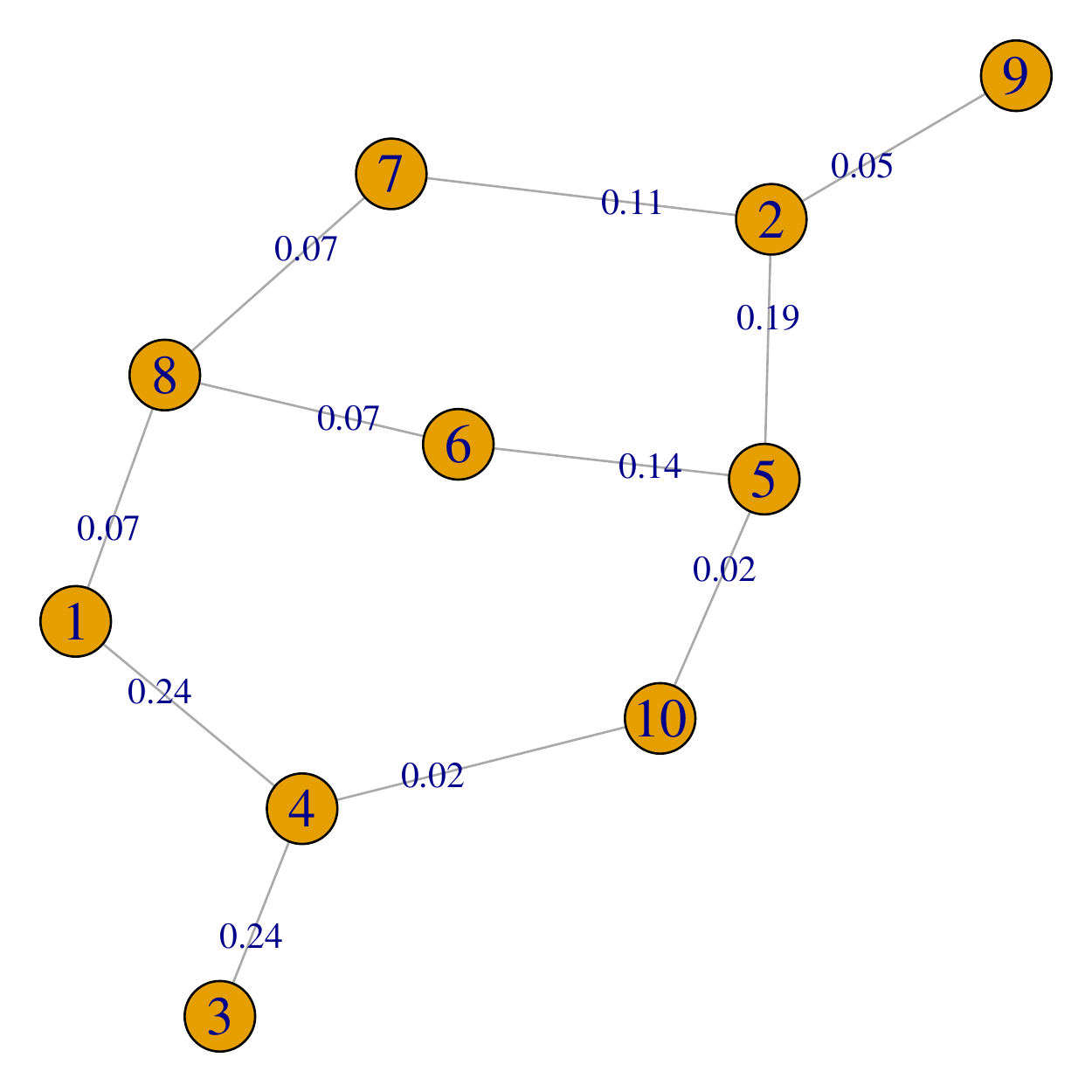}
        \label{Fig:simu3_1_structure}
    }
    \hspace{.05\textwidth}
    \subfigure[{\footnotesize{Case1: Error vs Iteration.}}]{
        \includegraphics[width=0.29\textwidth]{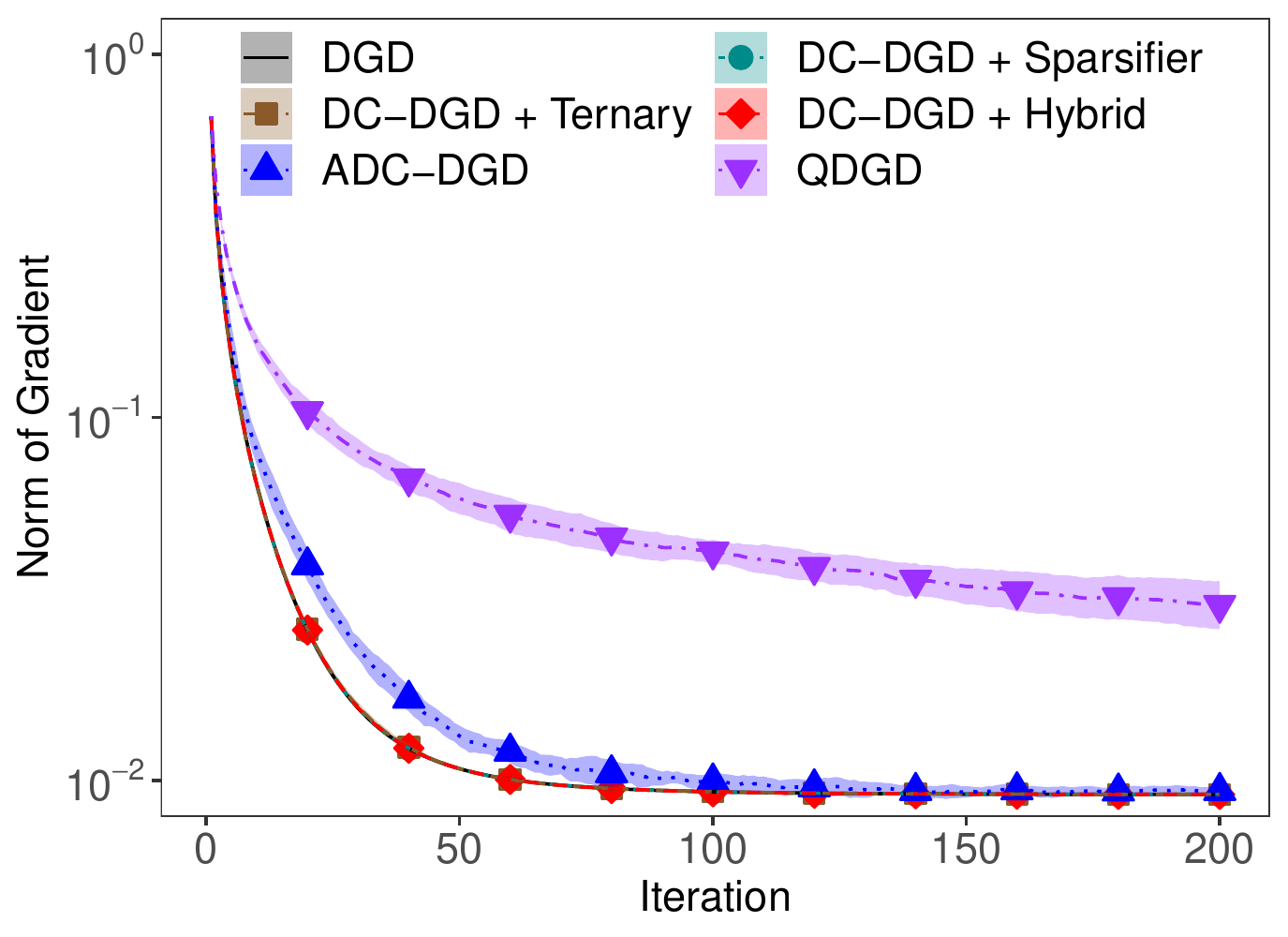}
        \label{Fig:simu3_1_iter}
    }
    \hspace{.05\textwidth}
    \subfigure[{\footnotesize{Case1: Error vs Comms. Cost.}}]{
        \includegraphics[width=0.29\textwidth]{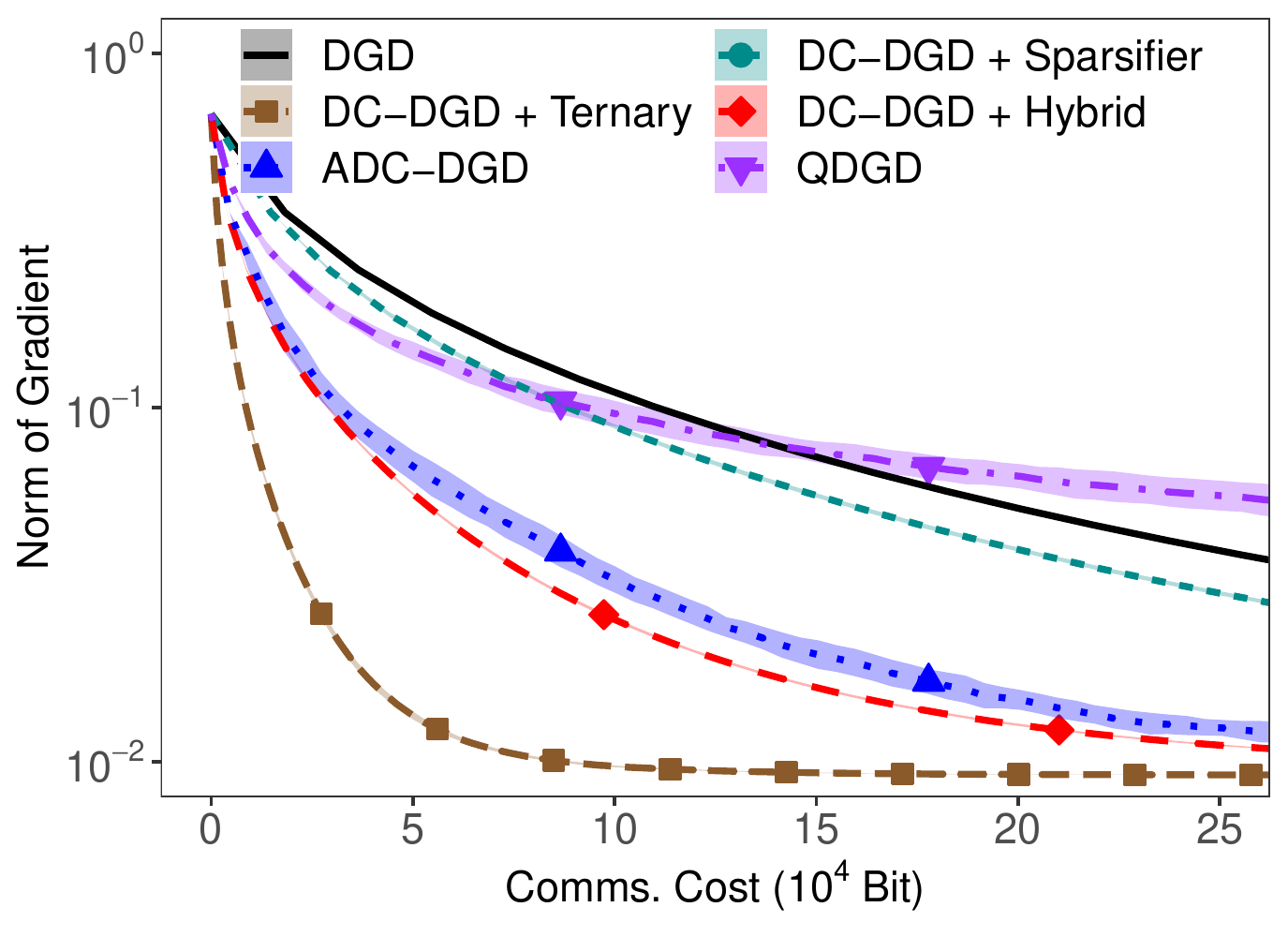}
        \label{Fig:simu3_1_cost}
    }\\
     \subfigure[{\footnotesize{Network Topology of Case2.}}]{
        \includegraphics[width=0.22\textwidth]{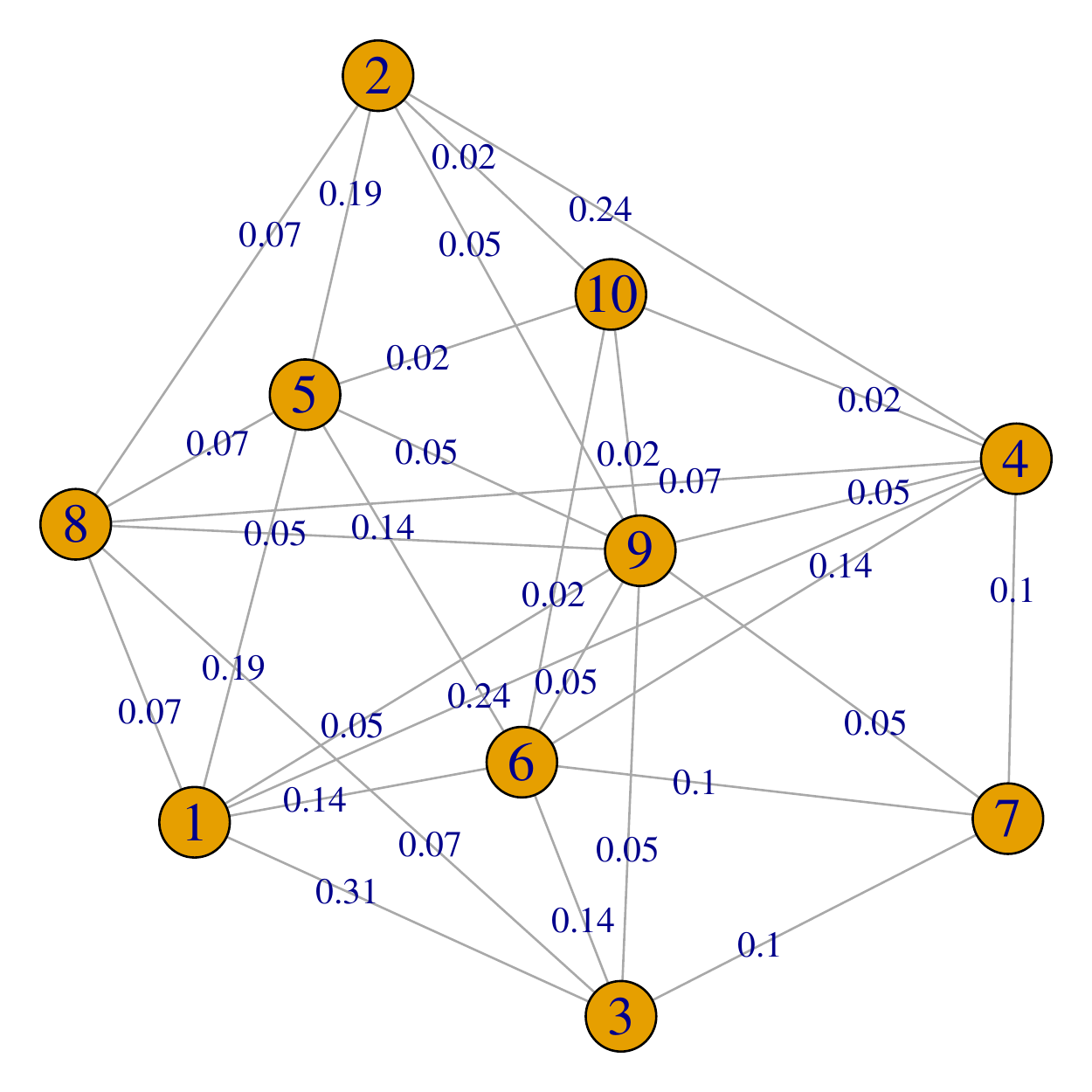}
        \label{Fig:simu3_2_network}
    }
    \hspace{.05\textwidth}
    \subfigure[{\footnotesize{Case2: Error vs Iteration.}}]{
        \includegraphics[width=0.29\textwidth]{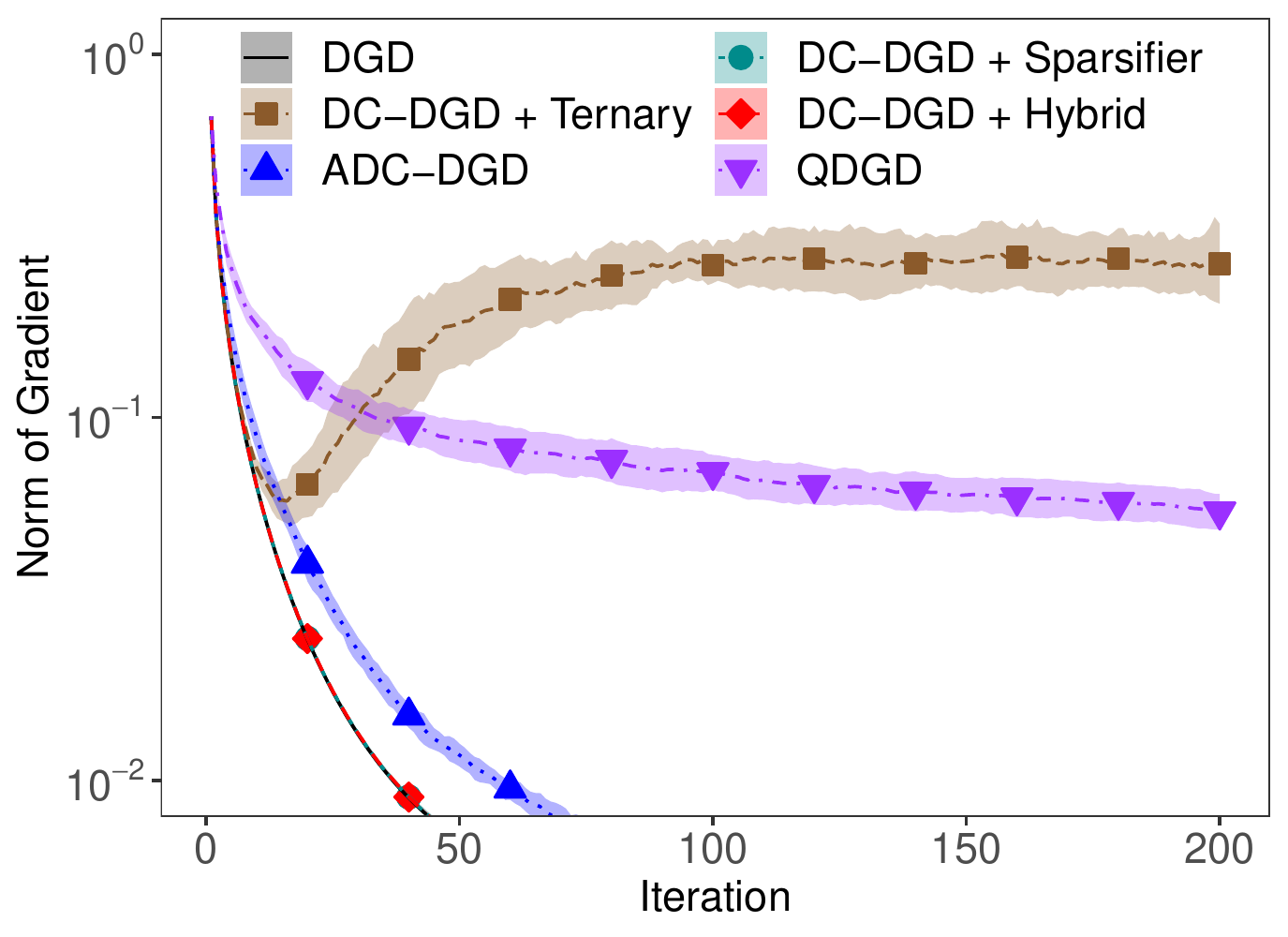}
        \label{Fig:simu3_2_iter}
    }
    \hspace{.05\textwidth}
    \subfigure[{\footnotesize{Case2: Error vs Comms. Cost.}}]{
        \includegraphics[width=0.29\textwidth]{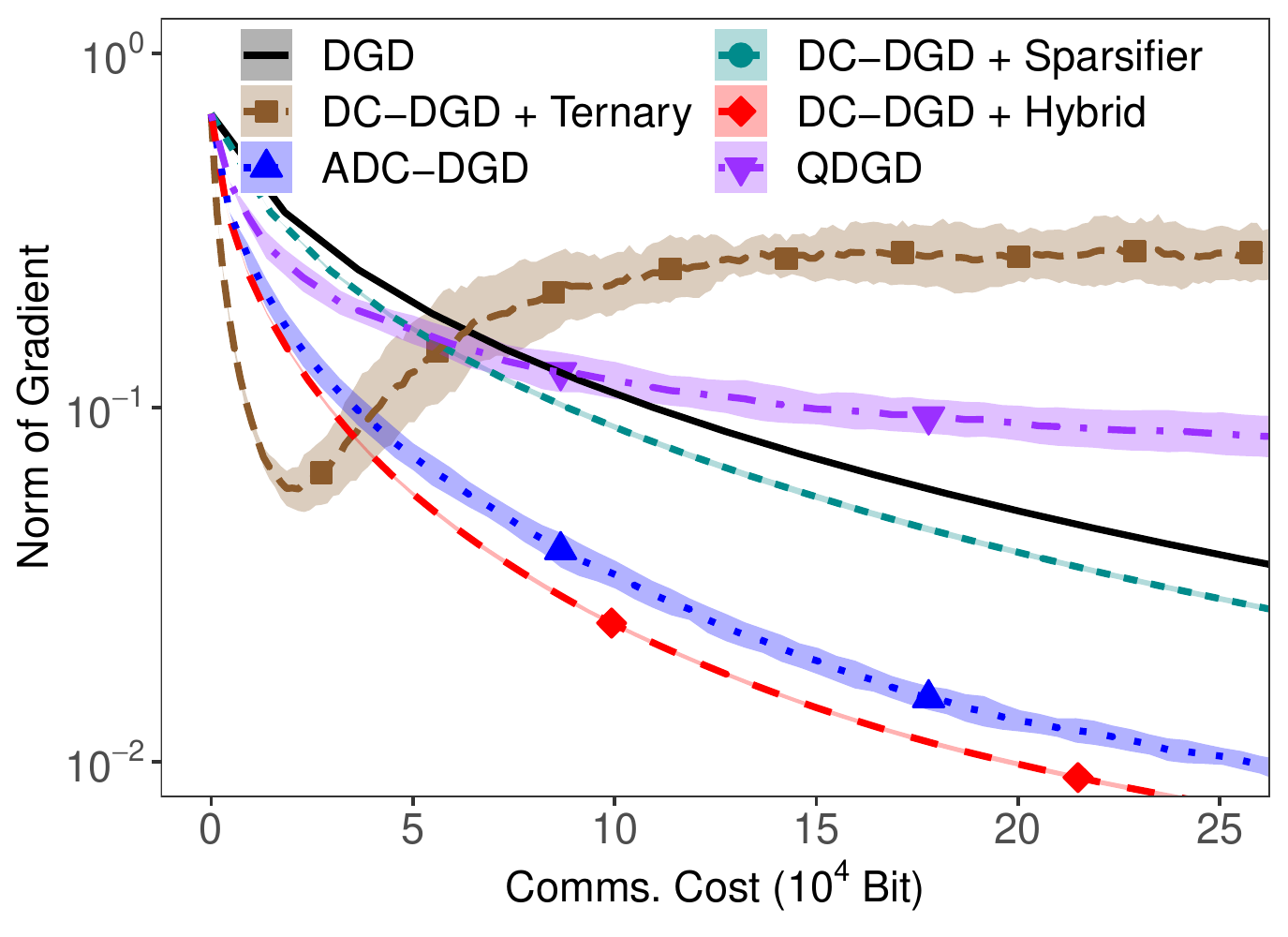}
        \label{Fig:simu3_2_cost}
    }
    \vspace{-.05in}
    \caption{(a) an (d) Two ten-node network examples. The consensus weights are shown on the corresponding edges. 
    (b) and (e) Convergence in terms of iterations; 
    (c) and (f) Convergence in terms of communication cost. 
    The curves are averaged over 10 trials and the shaded regions represent the standard deviation of results over random trials.}\label{Fig: Simu_3}
\vspace{-.2in}
\end{figure*}

\smallskip
{\bf 1) Convergence of DC-DGD:}
In this simulation, we adopt the sparsifier compression in Example~\ref{Ex: Sparse} and vary the probability parameter $p$ to induce different SNR threshold values. 
Consider a five-node circle network in Fig.~\ref{Fig:simu1_network} with the global objective function: $\min_\x f(\x) = f_1(\x)+f_2(\x)+f_3(\x)+f_4(\x)+f_5(\x)$, where 
\begin{align} \label{eqn_fi}
&f_i(\x) = 
\begin{cases}
\log(1+(\a_i^{\top} \x+b_i)^2/2), & \text{if~} i = 1,2;\\
(\a_i^{\top} \x - b_i)^2/2, &\text{if~} i = 3,4,5.
\end{cases}
\end{align}
In (\ref{eqn_fi}), the coefficients $\{\a_i,b_i\}_{i=1}^{5}$ are randomly generated from the standard Gaussian distribution.
Note that $f_{1}(\x)$ and $f_{2}(\x)$ are non-convex and the remaining are convex.
In our simulations, we use the two consensus matrices:
\begin{align*}
&\W_1 = 
\begin{bmatrix}
\vspace{.03in}
\frac{1}{5} & \frac{2}{5} & 0 & 0 & \frac{2}{5} \\
\vspace{.03in}
\frac{2}{5} & \frac{1}{5} & 0 & 0 & \frac{2}{5} \\
\vspace{.03in}
0 & \frac{2}{5} & \frac{1}{5} & \frac{2}{5} & 0 \\
\vspace{.03in}
0 & 0 & \frac{2}{5} & \frac{1}{5} & \frac{2}{5} \\
\vspace{.03in}
\frac{2}{5} & 0 & 0 & \frac{2}{5} & \frac{1}{5} 
\end{bmatrix}, \,\,
\W_2 =
\begin{bmatrix}
\vspace{.03in}
\frac{1}{2} & \frac{1}{4} & 0 & 0 & \frac{1}{4} \\
\vspace{.03in}
\frac{1}{4} & \frac{1}{2} & \frac{1}{4} & 0 & 0 \\
\vspace{.03in}
0 & \frac{1}{4} & \frac{1}{2} & \frac{1}{4} & 0 \\
\vspace{.03in}
0 & 0 & \frac{1}{4} & \frac{1}{2} & \frac{1}{4} \\
\vspace{.03in}
\frac{1}{4} & 0 & 0 & \frac{1}{4} & \frac{1}{2}
\end{bmatrix}.
\end{align*}
Note that $\lambda_N(\W_1)\!=\!-0.45$ and $\lambda_N(\W_2)\!=\!0.09.$
We compare the original DGD, the ADC-DGD\cite{zhang2018compressed}, and our DC-DGD algorithms. 
For DC-DGD, the sparsifier probability parameter $p$ is chosen from $\{0.3,0.5,0.8\}$.
Note that since ECD-PSGD and DCD-PSGD in\cite{tang2018communication} are using stochastic gradients and hence results are not directly comparable,
they are not included in the simulations.
 In ADC-DGD, we adopt the low-precision representation (see   \cite[Example~1]{reisizadeh2018quantized}) and choose the amplifying exponent $\gamma$ from $\{0.8,1.2\}.$ 
We use fixed step-size $0.1$ and repeat 50 independent trials for each setting.
The simulation results are presented in Figs.~\ref{Fig:simu1_1}--\ref{Fig:simu1_2}.

Fig.~\ref{Fig:simu1_1} illustrates the convergence of the three algorithms with $\W_1$. 
We can see that DC-DGD converges with $p=0.8$ but fails to converge with $p$ chosen from $\{0.3,0.5\}$.
This confirms our Theorem~\ref{Theorem: Theorem1}:
From Example~\ref{Ex: Sparse} and Theorem \ref{Theorem: Theorem1}, the lower bound of $p$ can be derived from $p/(p-1) > (1-\lambda_N(\W_1))/(1+\lambda_N(\W_1))$, which is $0.72$. 
Thus, choosing $p \in \{0.3, 0.5\}$ (i.e., $p< 0.72$) violates the convergence condition in Theorem~\ref{Theorem: Theorem1}. 
Moreover, we note that, with $p=0.8$, the convergence speed of the DC-DGD is {\em almost the same as the original DGD} (the black dashed line). 
Fig.~\ref{Fig:simu1_2} presents the convergence performance of these algorithms with $\W_2$. 
Following similar derivations, one can show that the lower bound of $p$ is $0.45$.
In this case, DC-DGD converges for $p=0.5$ and fails to converge for $p=0.3$, which confirms Theorem~\ref{Theorem: Theorem1} again.
In both cases, we can see that DC-DGD converges faster and has smaller variances than ADC-DGD.

\smallskip
{\bf 2) Compression Operator Comparison:}
Next, we compare three SNR-constrained compressors: the sparsifier, the ternary compressor, and our proposed hybrid compressor. 
We generate $20$ $d$-dimensional vectors independently from the multivariate Gaussian distribution $\mathcal{N}(\0, \I_d)$ with $d \in \{20,50\}$. 
We apply three operators on each vector, respectively, and conducted $100$ trials.
For any $\x$ and the compressed $C(\x)$, we evaluate: 1) Bias: $\|\mathbb{E}[C(\x)]-\x\|$; 2) Signal-to-Noise Ratio (SNR): $\|\x\|^2/\text{Var}[C(\x)]$; and 3) Communication Cost.
Here the SNR is corresponding to $\eta$ in Theorem \ref{Theorem: Theorem1}. 
The smaller bias and the larger the SNR (less noisy), the better the compressor.
To calculate the communication cost, we use $32$-bit floating numbers and $2$-bit ternary numbers. 
For the sparsifier operator, only one bit is used to represent value $0$. 
Note that SNR is controllable by adjusting $p$ in the sparsifier and our hybrid compressors.
To illustrate this advantage, we set the SNR lower bound as $0$ dB and $3$ dB. 
In both compressors, the parameters are optimized for the largest communication cost savings: 
For the $3$ dB SNR lower bound, we have $p=\frac{2}{3}$ for the sparsifier and $\eta = 2$ for the hybrid compressor; 
For the $0$ dB SNR lower bound, we have $p\!=\!\frac{1}{2}$ for the sparsifier and $\eta \!=\! 1$ for the hybrid compressor. 
Boxplots results are illustrated in Fig.~\ref{Fig: Simu_2}.

In Fig.~\ref{Fig:simu2_4} and \ref{Fig:simu2_1}, we can see that our hybrid compressor has the smallest bias, while the bias of the sparsifier increases as $p$ decreases. 
We can see from Fig.~\ref{Fig:simu2_5} and \ref{Fig:simu2_2} that our hybrid compressor can precisely make the SNR larger than the given bound, while the ternary operator fails to do so.
The communication costs are shown in Fig.~\ref{Fig:simu2_6} and \ref{Fig:simu2_3}.
Although the ternary compressor has the lowest cost, it cannot control its SNR. 
In contrast, our hybrid scheme achieves almost $50\%$ cost savings compared to the sparsifier scheme under all circumstances.

\smallskip
{\bf 3) Real-World Data Experiments:}
Lastly, we compare DC-DGD with the original DGD\cite{nedic2009distributed}, QDGD\cite{reisizadeh2018quantized}, ADC-DGD\cite{zhang2018compressed} in 10-node networks with real-world data. 
We consider a classification task on the Spambase dataset from UCI repository\cite{Dua:2019}. 
This dataset contains email spam data from $4601$ email messages and $57$ features. 
The data are evenly distributed to $10$ machines.
The local objective $f_i(\x)$ is a logistic regression problem with the non-convex regularizer\cite{wang2018stochastic}:
$-\frac{1}{n_i} \sum_{j=1}^{n_i} [y_{ij}\log (\frac{1}{1+\exp(-\x^\top\zeta_{ij})} ) + (1-y_{ij}) \log (\frac{\exp{(-\x^\top\zeta_{ij})}}{1+\exp{(-\x^\top\zeta_{ij})}})]+ \rho \sum_{i=1}^{d} \frac{x_i^2}{1+x_i^2}$,
where the label $y_{ij} \in \{0,1\},$ the feature $\zeta_{ij} \in \mathbb{R}^{57}$ and $\rho =0.1$ in our experiment.
For ADC-DGD and QDGD, floating numbers are randomly quantized to integers with the low-precision representation. 
In our DC-DGD, we test three compressors: the sparsifier, the ternary compressor, and our hybrid compressor. 
We use $32$ bits for the floating numbers, $8$ bits for integers (int8), and $2$ bits for ternary values. 
In addition, value $0$ is represented by $1$ bit in the sparsifier. 
We use two different network topologies as shown in Figs.~\ref{Fig:simu3_1_structure} and \ref{Fig:simu3_2_network}. 
For the first topology, $\beta = 0.98$ and $\lambda_N = 0.24$;
For the second topology, $\beta = 0.88$ and $\lambda_N = -0.37$.
The simulation results are shown in Fig.~\ref{Fig: Simu_3}.

%The first row in Fig.~\ref{Fig: Simu_3} shows the results of the first network topology while the second row shows those of the second structure. 
We can see that DC-DGD with the ternary compressor does not converge under the second topology.
%This confirms Theorem~\ref{Theorem: Theorem1}: 
This is because the SNR-threshold is {\em not} controllable under the ternary compressor. 
Thus, ternary compressor is not a safe choice in DGD-type algorithms.
Fig.~\ref{Fig:simu3_1_iter} and \ref{Fig:simu3_2_iter} illustrate the convergence rates of the algorithms.
We can see that the QDGD has the slowest convergence speed, which is followed by ADC-DGD.
Note that DC-DGD, when converged, has {\em almost the same speed as the original DGD}.
Fig.~\ref{Fig:simu3_1_cost} and \ref{Fig:simu3_2_cost} compare the communication cost of these algorithms.
In Fig.~\ref{Fig:simu3_1_cost}, we see that the ternary compressor has the lowest communication cost (approximately $10^5$ bits to achieve error $10^{-2}$).
However, ternary compressor does not work in the second network.
We can also see that DC-DGD with our hybrid compressor converges in both networks and has the {\em lowest communication cost} under the second network (approximately $2\times10^5$ bits to achieve error $10^{-2}$). 
In contrast, ADC-DGD costs $2.5\times10^5$ bits and other methods cost more than $2.5\times10^5$ bits.
Moreover, we note that our DC-DGD has smallest variance compared to ADC-DGD and QDGD (compare the shaded regions), which suggests that our DC-DGD is more stable.

\section{Conclusion} \label{Section: Conclusion}

In this paper, we designed and analyzed a new differential-coded compressed decentralized gradient descent (DC-DGD) algorithm for communication-efficient network-distributed optimization.
The key features of our DC-DGD algorithm include: 
i) DC-DGD works with general compression schemes that are only constrained by SNR (signal-to-noise ratio);
ii) By exchanging the differentials between successive iterations (hence the name differential-coded), the DC-DGD algorithm converges at the same $O(1/t)$ rate as the original DGD;
ii) DC-DGD enjoys the same low-complexity algorithmic structure as the original DGD algorithm and does not require additional mechanisms to tame compression noise thanks to its {\em self compression noise reduction effect}.
%On the theoretical side, we show that the compression noise under our algorithm structure is self-reduced as iteration and the optimal convergence rate of DC-DGD is proved as $O(T^{-2/3})$ under some mild conditions.
Based on the above theoretical insights, we proposed a new family of hybrid SNR-constrained compressors that integrate sparsifier and ternary operators.
We showed that our hybrid compressor has a controllable SNR-threshold and offers a systematic framework to minimize communication costs.
Moreover, by leveraging the special problem structure, we developed an efficient greedy algorithm to reduce the communication cost.
%We conducted extensive experiments to validate the efficacy of DC-DGD and the hybrid compressor.

%There are serveral {\color{red}{limitations}} in our work.
%First, we give an upper bound for determining the compression operator. 
%However, our bound is sufficient. 
%It will be interesting to explore the sufficient and necessary upper bound, with which the communication cost can be further reduced.
%Second, our algorithm can only get into an error ball of the stationary point with a fixed step-size and $O(1/T)$ rate.
%There exist serveral correction procedures (\cite{xu2017convergence,shi2015extra,li2017decentralized}), which can help DGD-based algorithms exactly reach the stationary point with a fixed step-size.
%These correction procedures can also be adopted for improve our communication-efficient DGD-based algorithm.
%Third, though our optimal hybrid operator can reduce more communication cost, it introduces extra computation with the greedy algorithm and without any approximation analysis.
%It will be of the great help to design and analyze a selection algorithm {\color{red}{with low complexity and exact solution.}}

\bibliographystyle{IEEEtran}
\bibliography{reference}

%\clearpage
%\appendix
%\input{Appendix/Proof}

\end{document}